\documentclass{ar-1col}
\usepackage{url}
\usepackage[numbers]{natbib}
\usepackage{rotating}
\usepackage{pdflscape}
\usepackage{amsmath,amssymb}
\usepackage[letterpaper]{geometry}
\bibpunct{(}{)}{;}{a}{}{,}

\usepackage{xparse}
\let\oldquote\quote
\let\endoldquote\endquote
\RenewDocumentEnvironment{quote}{om}
  {\oldquote}
  {\par\nobreak\smallskip
   \hfill(#2\IfValueT{#1}{~---~#1})\endoldquote 
   \addvspace{\bigskipamount}}

\makeatletter
\newsavebox\myboxA
\newsavebox\myboxB
\newlength\mylenA
\newcommand*\xoverline[2][0.75]{%
    \sbox{\myboxA}{$\m@th#2$}%
    \setbox\myboxB\null
    \ht\myboxB=\ht\myboxA%
    \dp\myboxB=\dp\myboxA%
    \wd\myboxB=#1\wd\myboxA
    \sbox\myboxB{$\m@th\overline{\copy\myboxB}$}
    \setlength\mylenA{\the\wd\myboxA}
    \addtolength\mylenA{-\the\wd\myboxB}%
    \ifdim\wd\myboxB<\wd\myboxA%
       \rlap{\hskip 0.5\mylenA\usebox\myboxB}{\usebox\myboxA}%
    \else
        \hskip -0.5\mylenA\rlap{\usebox\myboxA}{\hskip 0.5\mylenA\usebox\myboxB}%
    \fi}
\makeatother

\jname{Annual Reviews of Nuclear and Particle Science}
\jvol{AA}
\jyear{2018}
\doi{10.1146/((please add article doi))}

\begin{document}

\markboth{Frebel}{Tracing Heavy-Element Production with the Oldest Stars}
\title{From Nuclei to the Cosmos: \\Tracing Heavy-Element Production with the Oldest Stars}

\author{Anna Frebel$^1$
\affil{$^1$Department of Physics and Kavli Institute for
  Astrophysics and Space Research, Massachusetts Institute of
  Technology, Cambridge, MA 02139, USA; email: afrebel@mit.edu}}

\begin{abstract}

Understanding the origin of the elements has been a decades long pursuit, with many open questions still remaining. Old stars found in the Milky Way and its dwarf satellite galaxies can provide answers because they preserve clean elemental patterns of the nucleosynthesis processes that operated some 13 billion years ago. This enables the reconstruction of the chemical evolution of the elements. Here we focus on the astrophysical signatures of heavy neutron-capture elements made in the $s$-, $i$- and $r$-process found in old stars. A highlight is the recently discovered $r$-process galaxy Reticulum\,II that was apparently enriched by a neutron star merger. These results show that old stars in dwarf galaxies provide a novel means to constrain the astrophysical site of the $r$-process, ushering in much needed progress on this major outstanding question. This nuclear astrophysics work complements the many nuclear physics efforts into heavy-element formation, and aligns with recent results on the gravitational wave signature of a neutron star merger.  

\end{abstract}

\maketitle
\tableofcontents

\section{INTRODUCTION}

\begin{quote}[Stars and Atoms (1928), Lecture 1]{Sir Arthur Eddington}
I ask you to look both ways. For the road to a knowledge of the stars leads through the atom; and important knowledge of the atom has been reached through the stars.
\end{quote}

The discovery that stars are made primarily from hydrogen and helium \citep{payne25} led many scientists to realize that the study of the elements is intimately connected to understanding stars and their evolution  \citep{Burbidge57}. Nearly 100 years later, this work systematically continues, driven by stellar observations, nuclear physics experiments, and theoretical investigations to uncover the many interdisciplinary connections that will eventually lead to a more complete understanding of the cosmic origin of the chemical elements.

\subsection{From Nuclei to the Cosmos: The oldest stars as probes of the early universe}

This decade has seen a tremendous increase in observational surveys
and theoretical studies of our Milky Way Galaxy. In this ``Galactic Renaissance'', it is now widely recognized that a detailed knowledge of the assembly history of the Milky Way and its satellite dwarf galaxies provides important constraints on our understanding of a large variety of topics, ranging from nuclear physics and nuclear astrophysics, to stellar astronomy, to galaxy formation, and even observational cosmology. Studying the Milky Way, its stars and stellar populations, also enables a complementary approach to observing the most distant galaxies that formed less than a billion years after the Big Bang, e.g., with NASA's upcoming \textit{James Webb Space Telescope}. Ancient, undisturbed, 13 billion-year-old stars are now regularly found in our own Galaxy, and provide the opportunity to look back in time -- right into the era of the very first stars -- and elucidate the nucleosynthetic pathways to the formation of the elements, and the beginning of the chemical evolution that proceeds to this day.

The early history of the 13.8-billion-year-old Universe is
encoded in the chemical compositions of ancient, low-mass [\textit{M} $\le$
0.8 solar mass (M$_{\odot}$)] stars. Even today, the atmospheres
of these still-shining stars reflect the composition of their
birth gas clouds at the time and place of their formation some 13
billion years ago. They formed from the nearly pristine gas left over
after the Big Bang, when the Universe was just starting to be enriched in elements
heavier than hydrogen and helium. Thus, a low abundance of
elements heavier than hydrogen and helium, such as iron, observed in
a star indicates a very early formation time. It is these stars
that are of greatest value for \textit{stellar archaeology} and
\textit{dwarf galaxy archaeology}, because they provide a
remarkably powerful tool for investigating the physical and
chemical conditions of the early Universe. For example, the oldest
stars enable us to isolate clean signatures from various
nucleosynthesis processes that occurred prior to their formation.
In this way, they uniquely contribute to {nuclear
astrophysics} by providing information about elements that
cannot be made or studied with nuclear physics experiments. From
an astrophysics standpoint, these aspects of nuclear astrophysics
explain how chemical enrichment proceeded in the cosmos, how the
content of elements evolved with time, and how the presence of
elements influenced early star and galaxy formation. In addition,
studying these old stars enables one to connect the present state
of the Galaxy with its long assembly history involving the
accretion of neighboring smaller galaxies. These efforts are part
of {near-field cosmology}, which aims to interpret the
chemical abundances of the oldest stars from a broader,
cosmological perspective.

\begin{figure}[!t]
\includegraphics[width=5.9in]{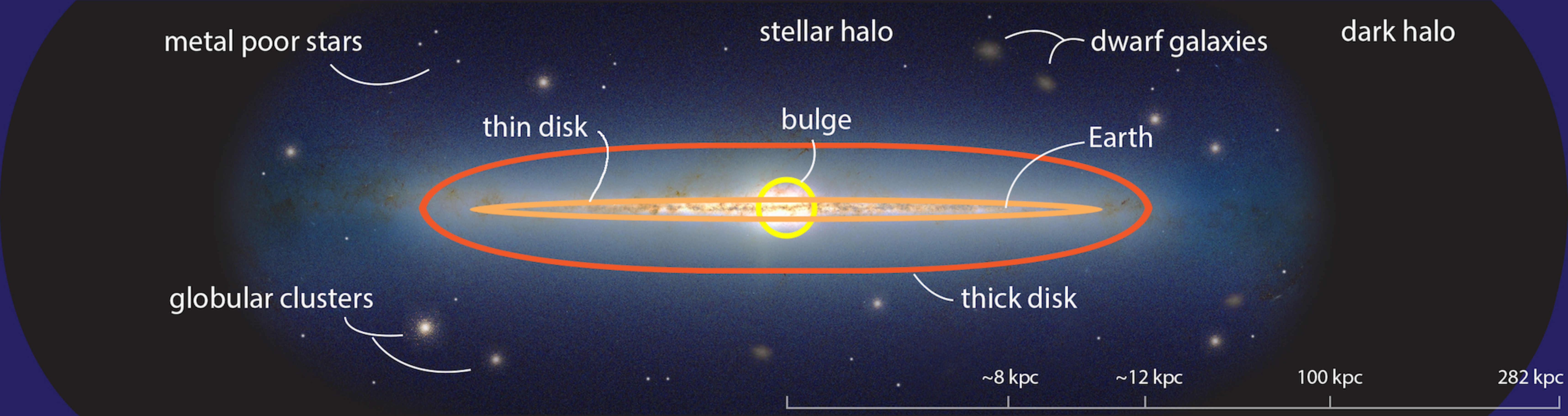}
\caption{\label{mw_schematic} Structure and stellar populations of the Milky Way. Stellar archaeology is based on old, metal-poor halo stars. Dwarf galaxy archaeology utilizes stars in satellite dwarf galaxies that orbit the Milky Way. Figure courtesy of K. Brauer.}
\end{figure}

In this context, I briefly review the structure and stellar
populations of the Galaxy \citep{bland_haw16} as illustrated in  \textbf{Figure \ref{mw_schematic}}. The Milky Way is a spiral disk galaxy
containing a few hundred billion stars. It has a total
dark matter mass of $(1.3 \pm 0.3)\times10^{12}$\,M$_{\odot}$ and a
total stellar mass of $(5\pm1)\times10^{10}$\,M$_{\odot}$. The Milky Way and Andromeda, its slightly more massive sister galaxy, are by far the most massive galaxies in the Local Group. They
are surrounded by approximately 50 known much smaller dwarf galaxies
(with dark matter masses of $10^{6}$--$10^{10}$\,M$_{\odot}$). The
Galactic disk, which has a radius of 10--15\,kpc, consists of
several components: a thin disk with a vertical scale height of
$300\pm50$\,pc, a thick disk with a scale height of $900\pm180$\,pc,
and an even more vertically extended so-called metal-weak thick
disk. The central $\sim3$\,kpc region of the Galaxy, the somewhat
X-shaped bulge, is a very dense and luminous region of ongoing
star formation that surrounds the Milky Way's central supermassive
black hole of ($4.2\pm0.2)\times10^{6}$\,M$_\odot$. Both the disk and, especially,
the bulge are old structures, but they contain predominantly younger
stars with near-solar abundances, although the bulge might contain
a small number of the oldest stars left over from the Milky Way's
earliest formation phase. For reference, the Solar System is
located $8.2\pm0.1$\,kpc from the Galactic Center. The disk(s) and
bulge are enveloped by the so-called stellar halo, which is of much
lower stellar density, and likely extends well beyond 100\,kpc,
where the farthest known stars are currently found. Also, the
virial radius that includes the dark matter halo of the Galaxy is
$280\pm30$\,kpc. The stellar halo with
(4--$7)\times10^{8}$\,M$_{\odot}$ hosts primarily older stars,
globular star clusters, and those satellite dwarf galaxies that
orbit the Milky Way. Debris and tidally ripped-apart dwarf
galaxies form halo substructures with a combined
(2--$3)\times10^{8}$\,M$_{\odot}$ and are responsible for
extended star streams that span across the sky. Examples include the
Sagittarius Stream, which originates in the Sagittarius dwarf
galaxy and has left stellar debris throughout the halo, and the
Monoceros Stream, which wraps twice around the Galaxy near the
disk plane.

Stellar streams are examples of how stars from dwarf galaxies can become mixed into the Milky Way. Once they are in the halo and several billion years have passed, is (near) impossible to correctly assess their origin outside of the Galaxy. The same holds true for the oldest halo stars. They are likely older than the Milky Way itself and thus must have formed in some of the earliest star-forming galaxies, not unlike those dwarf galaxies that survived to still orbit the Milky Way today. These early galaxies were later absorbed by the Galaxy, as part if its own hierarchical growth, spilling all their stars into the stellar halo. The consequence is that, when we identify and work with these stars, we can only speculate about their origins and birth environments. On the contrary, this highlights the power of searching for old stars in the satellite dwarf galaxies. These stars are still located within their natal system. It affords us the opportunity to assess their environment to derive firm conclusions about element-production events, how elemental yields were dispersed through the galaxy, and how the yields eventually got incorporated into the dwarf-galaxy stars we observe. Studying stars and stellar populations in the satellite dwarf galaxies therefore provides an excellent complementary approach to working with halo stars. As will be outlined in Section~\ref{ret}, research based on the compositions of individual stars in dwarf galaxies recently led to a breakthrough in understanding the astrophysical site of the heaviest elements.

\subsection{Review aims and further reading}\label{reading}

This review provides a compact overview of the recent progress regarding the origin and early evolution of the heavy elements that are made during the rapid neutron-capture process, as told by an astronomer. We aim to highlight links between what is studied by nuclear physicists (nuclear properties of matter), and what astronomers observe  (stars with chemical abundance signatures that are the end result of various nucleosynthesis processes). New experimental nuclear physics facilities, such as the Facility for Rare Isotope Beams (FRIB), will investigate neutron-rich nuclei far away from stability, which promises to yield an improved understanding of heavy-element production. Observations of the oldest stars in the Milky Way and its satellite dwarf galaxies provide complementary insights. They preserve a fossil nucleosynthesis record of astrophysical events of element production, providing valuable details concerning the nucleosynthesis processes involved and their associated astrophysical sites of operation.

Focusing here on the topic of neutron-capture elements implies that much of the related information on old stars, stellar archaeology, dwarf galaxy archaeology, near-field cosmology, and even nuclear astrophysics can unfortunately not be covered. We instead refer the interested reader to the following reviews: 

\vspace{-0.2cm}
\begin{enumerate}

\item[--] A detailed overview of near-field cosmology, stellar archaeology and dwarf galaxy archaeology with the oldest stars  \citep{fn15}; additional introductory material on the subject is covered by \citep{psss} 

\item[--] Early progress including the history of the search for old stars in the Galaxy \citep{beers&christlieb05}

\item[--] Observations of neutron-capture elements in old stars and their interpretation  \citep{sneden08,jacobson13} 

\item[--] Reviews on neutron star mergers and associated heavy element production \citep{fernandez16,thielemann17}

\item[--] Understanding the Galaxy in a cosmological context \citep{bland_haw16} and its formation when interpreting the chemical abundances of old stars \citep{blandhawthorn_freeman}
\end{enumerate}
\vspace{-0.2cm}

\section{CHEMICAL ABUNDANCE MEASUREMENTS OF OLD METAL-POOR STARS}\label{sec2}

We first summarize some technical details associated with the spectroscopic analysis of old stars required to extract their chemical element signatures. Typically, these signatures are then interpreted by comparing them with theoretical predictions of the elemental yields of various nucleosynthesis processes and/or astrophysical sources and sites. We then highlight the characteristics of the chemically most primitive stars, i.e., the most iron-poor stars, to show how stellar abundances are used to learn about the nature and properties of the first stars.

\subsection{Astronomy jargon and classifications of metal-poor stars}\label{jargon}

\begin{figure}
\centering
\begin{minipage}{.4\textwidth}
  \centering
  \includegraphics[width=2.35in]{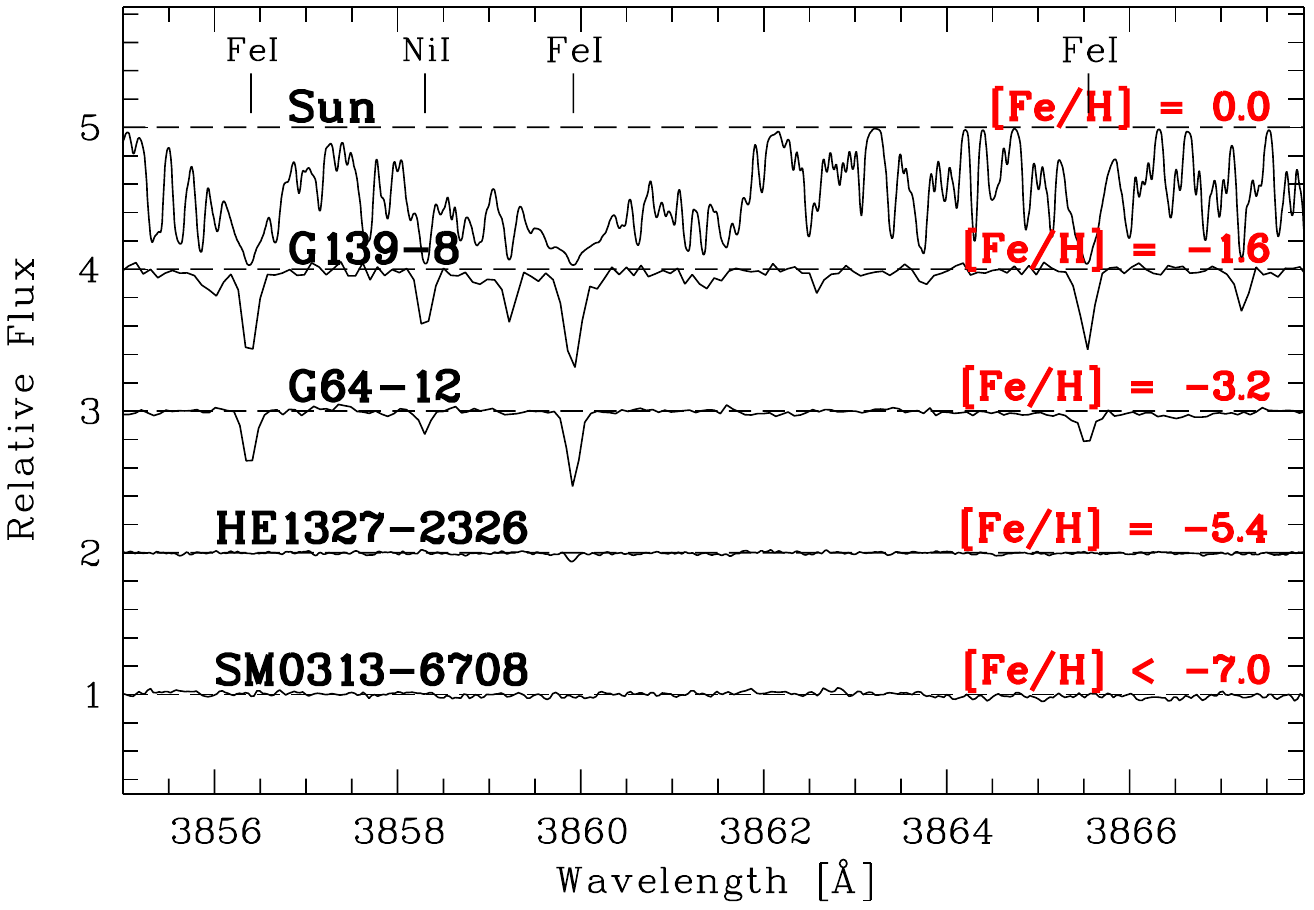}
\end{minipage}%
\begin{minipage}{.6\textwidth}
  \centering
  \includegraphics[width=3.35in]{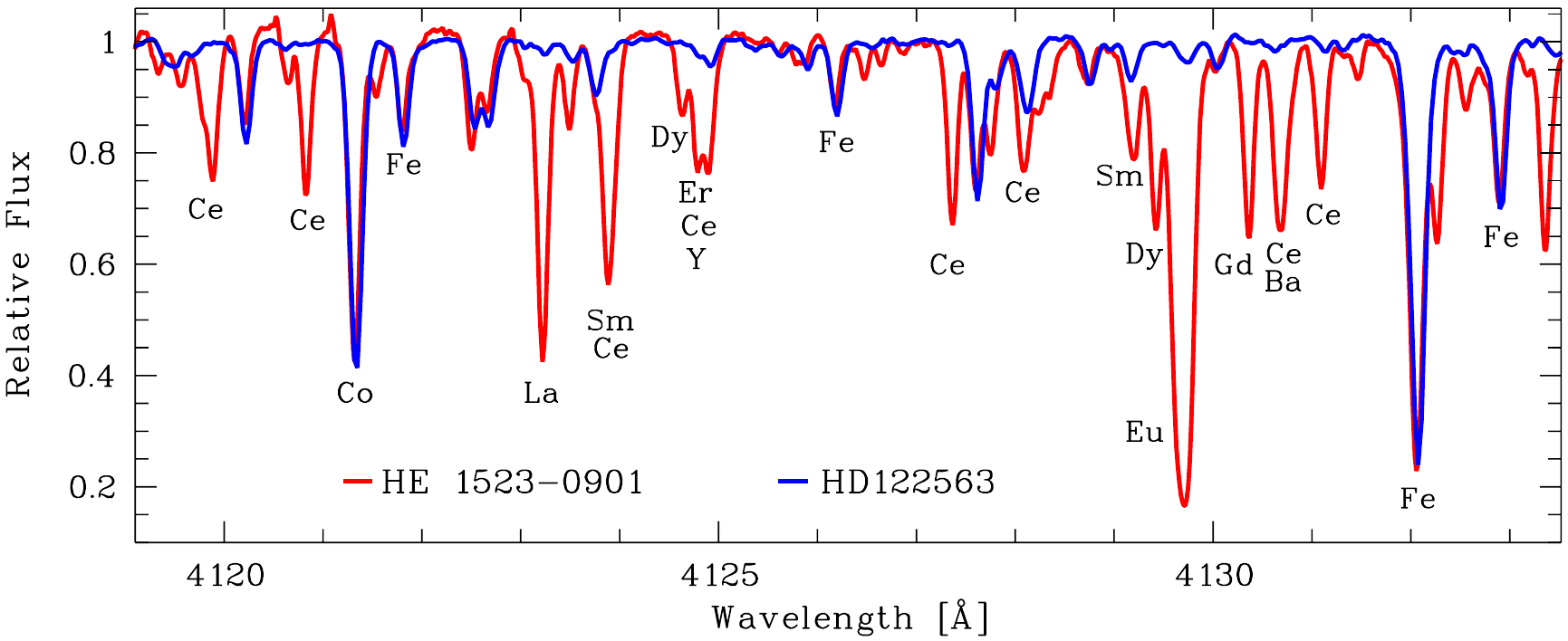}
\end{minipage}
\caption{\label{comp} Left: Spectral comparison of stars with different metallicities, as seen by decreasing line strengths. The bottom star has no iron lines detected \citep{keller14}. Several absorption lines are marked. Figure adapted from \citet{frebel10_AN}. Right: Spectral comparison around the Eu\,II line at 412.9\,nm of the $r$-process deficient star HD122563 and the $r$-process-enhanced star HE~1523$-$0901. Both stars have similar temperatures and metallicities. Figure from \citet{frebel08}.}
\end{figure}

In astronomy, the term ``metals'' collectively refers to \textit{all} elements heavier than hydrogen and helium. Most stars are made from roughly 71\% hydrogen, 27\% helium, and up to a few percent of metals, i.e., the other elements combined. However, the amount of metals in a star, also called ``metallicity'', also depends on its formation time and birth place. After the Big Bang, the universe consisted of just hydrogen (75\%) and helium (25\%) and trace amounts of lithium. The very first stars that emerged a few hundred million years later had masses of order 100\,$M_\odot$ \citep{bromm02}, and likely spanning $\sim10$ to several hundred $M_\odot$. Metal-free gas lacks sufficient coolants in the form of metals and/or dust grains that radiate away the thermal energy necessary for the gas to readily undergo gravitational collapse. Basic star formation models thus predict the formation of exclusively massive and short-lived stars, whereas stars with sub-solar masses can only form once metals in the gas lead to more efficient fragmentation. Metals were indeed forged in the hot cores of the first stars in a series of fusion processes, leading to elements up to and including iron. Upon the explosions of the stars, the newly created material is dispersed into space, enriching the surrounding gas from which the next generation of stars forms. This implies that the chemical evolution is the successive build up of metals with cosmic time, illustrated in \textbf{Figure~\ref{comp}}. It began with the element enrichments by the first exploding stars, and has progressed to a state where stars that are currently forming contain 2-4\% of metals \citep{feltzing01}. Consequently, stars that formed before the Sun was born 4.6\,billion years ago generally have a lower metal content. Since the Sun is used as a reference, these older stars are called ``metal-poor'' (compared to the Sun). Stars similar to the Sun (which is most stars) are referred to as ``metal-rich''. 

A handful of stars are known that are so metal-poor that they are believed to be members of the second-generation of stars (i.e., the first long-lived low-mass stars) to have formed after the Big Bang. The amount of metals is so small that all heavy elements present in those stars could have been produced by a single progenitor star -- one of the very first stars. For reference, the Sun formed from a gas cloud that was enriched by many nucleosynthesis processes and sites over the course of $\sim9$ billion years. The difference with respect to second-generation stars would be the equivalent of order 1000 subsequently exploding massive stars enriching only the local gas. In reality, the process of chemical enrichment of a certain region is of course more complex, and requires detailed modeling of all associated chemical, dynamical and other evolutionary processes.

To describe the amount of metals in a star (i.e., its metallicity), one usually takes the
abundance of iron as a proxy. This is part
convenience, part necessity. It is not possible to determine the
abundances of {all} elements from stellar spectra and add
them up to obtain the total metal content. Instead, iron has
numerous spectral absorption lines, and is thus relatively easy to
measure. Chemical abundances of any two elements are then defined
relative to the respective abundances in the Sun as \mbox{[A/B]}$
= \log(N_{\rm A}/N_{\rm B}) - \log(N_{\rm A}/N_{\rm B})_\odot$ where
$N_{\rm {A}}$ ($N_{\rm {B}}$) is the number of atoms of element A (B). By definition, the Sun has a metallicity
of $\mbox{[Fe/H]}=0.0$. An extremely metal-poor star would have
$\mbox{[Fe/H]}=-3.0$, which corresponds to one thousandth of the solar
iron abundance. Super-metal-rich stars have positive [Fe/H]
values. Accordingly, the iron abundances are often thought to
indicate a rough formation time, even though little is known
about the complex age--metallicity relation \citep{starkenburg17}.
Stars with vastly different metal content have been discovered to
date, spanning a total of at least {eight orders of
magnitude}, from $\mbox{[Fe/H]}\sim+0.7$ to $\mbox{[Fe/H]}<-7.3$
 \citep{ness16, keller14}. This is shown in \textbf{Figure \ref{comp}}. In the latter
case, the iron abundance is so low that only an upper limit could
be determined. 

Numerous studies in the past three decades have
validated that the iron abundance is indeed a good proxy for the overall
metal content of a star. However, at the lowest metallicities,
specifically $\mbox{[Fe/H]}<-3.5$, this fortuitous relation breaks
down. A significant fraction of the stars show, for example, large
enhancements of carbon over iron compared with the Sun \citep{fn15},
significantly increasing the total metallicity relative to just the iron abundance. Thus, it is
important to refer to those stars as the most iron-poor instead of the most metal-poor.

\begin{table}[!t]
\tabcolsep7.5pt
\caption{\label{categories} Classes and Signatures of Metal-Poor Stars}
\begin{center}
\begin{tabular}{|l|c|c|c|c@{}}
\hline
Description & Definition & Abbreviation \\
\hline
Population\,III stars & Postulated first stars, formed from metal-free gas& Pop\,III\\
Population\,II stars & Old (halo) stars formed from low-metallicity gas & Pop\,II\\
Population\,I stars & Young (disk) metal-rich stars & Pop\,I\\ \hline

  Super metal-rich    & $\mbox{[Fe/H]} >0.0$  &   MR \\
  Solar               & $\mbox{[Fe/H]} =0.0$  &      \\
  Metal-poor          & $\mbox{[Fe/H]}<-1.0$  &   MP \\
  Very metal-poor     & $\mbox{[Fe/H]}<-2.0$  &   VMP \\
  Extremely metal-poor& $\mbox{[Fe/H]}<-3.0$  &   EMP \\
  Ultra metal-poor    & $\mbox{[Fe/H]}<-4.0$  &   UMP \\
  Hyper metal-poor    & $\mbox{[Fe/H]}<-5.0$  &   HMP \\
  Mega metal-poor     & $\mbox{[Fe/H]}<-6.0$  &   MMP \\
  Septa metal-poor    & $\mbox{[Fe/H]}<-7.0$  &   SMP \\
  Octa metal-poor     & $\mbox{[Fe/H]}<-8.0$  &   OMP \\
  Giga metal-poor     & $\mbox{[Fe/H]}<-9.0$  &   GMP \\
  Ridiculously metal-poor& $\mbox{[Fe/H]}<-10.0$ &   RMP \\
\hline\hline

Signature               & Metal-poor stars with neutron-capture element patterns      &Abbreviation\\\hline

Main $r$-process        &  $0.3 \le \mbox{[Eu/Fe]} \le +1.0$ and $\mbox{[Ba/Eu]} < 0.0$ & $r$-I  \\
                        &  $\mbox{[Eu/Fe]} > +1.0$ and $\mbox{[Ba/Eu]} < 0.0$           & $r$-II \\\hline

Limited $r$-process$^{a}$ & $\mbox{[Eu/Fe]} < 0.3$, $\mbox{[Sr/Ba]} > 0.5$, and  $\mbox{[Sr/Eu]} > 0.0$ & $r_{lim}$ \\\hline

$s$-process:            &  $\mbox{[Ba/Fe]} > +1.0$, $\mbox{[Ba/Eu]} > +0.5$; also $\mbox{[Ba/Pb]} > -1.5$        & $s$\\\hline

$r$- and $s$-process    & $0.0 < \mbox{[Ba/Eu]} < +0.5$ and $-1.0<\mbox{[Ba/Pb]}<-0.5$$^{b}$ &$r+s$\\\hline

$i$-process             &   $0.0 < \mbox{[La/Eu]} < +0.6$ and $\mbox{[Hf/Ir]}\sim1.0$$^{c}$ & $i$\\\hline\hline

Signature               & Metal-poor stars with other element characteristics         &Abbreviation\\\hline

Neutron-capture-\textit{normal} &  $\mbox{[Ba/Fe]} < 0$                               & no \\\hline

Carbon-enhancement     & $\mbox{[C/Fe]} > +0.7$, for $\log (L/L_{\odot} ) \le 2.3$    &  CEMP$^{d}$\\
                       & $\mbox{[C/Fe]} \ge (+3.0 - \log(L/L_{\odot} ))$, \mbox{for}
                         $\log(L/L_{\odot}) > 2.3$$^{e}$&   \\\hline\hline

$\alpha$-element enhancement   &  $\mbox{[Mg, Si, Ca, Ti/Fe]} \sim +0.4$  & $\alpha$-enhanced \\\hline

\end{tabular}
\end{center}
\begin{tabnote}
$^{a}$ Also referred to as the Light Element Primary Process (LEPP; \citealt{Travaglio04}).\\
$^{b}$ Based on just one known CEMP-$r+s$ star \citep{gull18}; may require future adjustments. \\
$^{c}$ If [Hf/Ir] is not available, use only [La/Eu]. Definition may require future adjustments (F. Herwig, personal comm.). \\
$^{d}$ Carbon-Enhanced Metal-Poor stars; the CEMP star definitions are from \citet{aoki07}. $s$- and $i$-process-enhanced stars are always CEMP stars; $r$-process-enhanced stars may or may not be CEMP stars; there is also the class of CEMP-no stars. \\
$^{e}$  Carbon corrections as function of luminosity can also be obtained from \citet{placco14} \\
\end{tabnote}
\end{table}

To easily differentiate various classes of
low-metallicity stars,  \citet{beers&christlieb05} introduced categories that classify their metallicity and
chemical signatures. {Table~\ref{categories}} presents an updated
version, with some modifications and additions from
\citet{psss}. The table first lists stellar-population classifications
that are primarily of a historic nature but are still widely used.
Next are the different metallicity classes, now extended to
$\mbox{[Fe/H]}<-10.0$, following suggestions by T. Beers
and I. Roederer (personal communications), followed by the
stellar chemical signatures associated with enhancements in
neutron-capture elements (i.e., those elements heavier than iron).
The abundance criteria listed offer a quick classification;
matching many more elemental abundances with the respective full
patterns will confirm the nature of the star. \textbf{Figure~\ref{comp}} shows
spectra of a regular metal-poor halo star and one enriched in
heavy $r$-process elements. 

The most significant change from the
original classifications is the addition to Table~1 of intermediate neutron-capture process ($i$-process)-enhanced
stars, which are further discussed in Section~\ref{iprocstars}.
Other neutron-capture-rich signatures are discussed in
Sections~\ref{sprocstars} and \ref{rprocstars}, followed by the
definition of carbon-enhanced metal-poor (CEMP) stars \citep{aoki07}. More information on this prominent group of metal-poor
stars can be found elsewhere \citep{aoki07,placco14,Yoon16}. For
completeness, I\ also include the commonly used criterion for \mbox{$\alpha$-enhancement}.

\subsection{Element Abundance Analysis Techniques}

The main goal of stellar spectroscopy is usually to measure the detailed chemical composition of the star (another reason is to obtain a star's radial velocity for kinematic analyses). This technique, and the telescopes used, is further described by \citep{psss} and \citep{fn15}. We only summarize key points here for completeness. A high-resolution spectrum (with resolving power $R=\lambda/\Delta \lambda > 20,000$) covering visible wavelengths, ideally from $\sim$350 to $\sim700$\,nm is best suitable for a detailed elemental abundance analysis, although spectra with low and medium-resolution ($1,000<R<6,000$) also deliver abundances of a few key elements. Spectra obtained as part of large surveys often have low-resolution, i.e. SDSS. 
Examples of spectra at different resolution can be found in \citet{psss}. Near-infrared spectra can also be used to obtain a number of elements, as demonstrated in the APOGEE survey \citep{garciaperez16}. Near-UV spectra are of considerable interest also, especially for obtaining abundances of heavy neutron-capture elements with transitions only found there, but such spectra are notoriously difficult to obtain (at present) for any but the very brightest stars (12 $< V < $9). This is due to the combination of the fact that stars of interest have low UV flux and the \textit{Hubble Space Telescope}'s mirror size of only 2.4\,m.

Elemental abundances are obtained from the measured strengths of the spectral absorption lines of neutral and singly ionized species in the outer stellar atmospheres. This implies that abundances will be surface abundances, but they still reflect the overall stellar abundances well, because of the unevolved nature and simple structure of the stars considered here. Then, abundances can be calculated from the line strengths using a model atmosphere that approximates the physical conditions in the outer layers of the star where the absorption occurs. One more ingredient is required, namely the so-called stellar parameters, effective temperature (T$_{\rm{eff}}$), surface gravity ($\log g$), chemical composition ([Fe/H]), and microturbulence (v$_{\rm{mic}}$). These parameters describe the atmosphere of a given star reasonably well, facilitating the choice for a model atmosphere to ultimately obtain the elemental abundances. Alternatively, synthetic spectra of known abundances can be computed and compared with the observed spectrum, as is often done in the case of molecular bands, or complex and blended absorption features, as is the case for many lines of neutron-capture elements. A detailed discussion on how to determine stellar parameters and chemical abundances using various techniques can be found in \citet{psss}.

We still note that the vast majority of stellar abundances are derived using simplified one-dimensional model atmospheres that assume local thermodynamic equilibrium (LTE) for the formation of all absorption lines in the outer stellar atmosphere. Considerable effort is ongoing in the community to change this to a framework based on non-local thermodynamic equilibrium (NLTE) assumptions and three-dimensional (3D) hydrodynamical modeling \citep{tremblay13, sitnova16, amarsi16}. Recent progress using quantum mechanical atomic inputs for hydrogen collisions to determine stellar iron abundances under NLTE conditions is described in \citet{ezzeddine16}. This is particularly relevant, since NLTE effects have been shown to alter abundances derived from numerous and commonly used Fe\,I lines much more significantly than those of Fe\,II lines \citep{asplund_araa,mashonkina11,lind12,bergemann12}. With the corrections being positive, the metallicity of the stars increases, and they become more metal-rich: only 0.25\,dex for extremely metal-poor stars with $\mbox{[Fe/H]}\sim-3.0$ (compared to typical Fe uncertainties of 0.1-0.2\,dex) but 0.7\,dex for the most iron-poor stars $\mbox{[Fe/H]}\sim-6.0$ which is significantly above any observational uncertainty. Computationally intensive 3D-LTE (and even some 3D-NLTE) stellar model atmospheres remain difficult to use for large samples of stars. Recent works also show significant deviations especially for the most metal-poor stars and for elements derived from molecular species, such as CH \citep{nordlander17,gallagher16}

Despite those challenges, abundances of the light fusion elements up to and including iron are available \citep{abohalima17} for more than a thousand metal-poor stars. The most iron-poor stars \citep{HE1327_Nature,caffau11, aguado18} are the easily accessible \textit{local equivalent} of the high-redshift universe because they probe the very first chemical enrichment events that set in motion the chemical evolution. Hence, these stars are used to reconstruct the properties of Population\,III first stars, their supernova explosions, and associated nucleosynthesis processes. This way, early metal and gas mixing processes, and assembly processes of galaxies can constrained, aiding theoretical works on key questions related to the emergence of the first stars and galaxies from the primordial universe, and how the chemical elements produced in stars and supernova explosions were recycled through galaxies and eventually into planets and life. Given space constraints, the reader is referred to \citep{fn15} for an in-depth discussion on the role of these lighter elements and the latest results.

\section{SIGNATURES OF NEUTRON-CAPTURE ELEMENTS IN METAL-POOR HALO STARS}\label{sec3}

With the exception of hydrogen and helium, elements up to and including zinc (with atomic number $Z\le30$) are created through fusion processes in the hot cores in stars during stellar evolution, and/or in explosive processes during their supernova explosions if the star was massive enough to explode (i.e., M $> 8$\,M$_{\odot}$). The remaining heavier elements are built-up in neutron-capture processes, when the astrophysical environment provides seed nuclei (e.g., Fe) and a certain flux of free neutrons. During rapid ($r$-) neutron-capture, i.e. the $r$-process, the neutron-capture rate exceeds that of the $\beta$-decay, and large numbers of neutrons can be captured in 1-2\,seconds.  This results in heavy neutron-rich nuclei that then decay back to stability to form the heaviest elements in the periodic table, from Sr ($Z=38$) to U ($Z=92$). About half of all stable isotopes of elements heavier than zinc are synthesized this way \citep{arlandini1999, sneden08}. The other half of the isotopes is build up over much longer timescales, from slow ($s$-) neutron-capture where the capture rate is less than the $\beta$-decay rate of the unstable isotopes. The postulated intermediate ($i$-) process is somewhat similar to the $s$-process, but requires significantly higher neutron densities. In the following, we describe the $s$-, $i$-, and $r$-processes, and highlight discoveries of metal-poor stars with their respective elemental signatures. 

\subsection{The astrophysical signature of the $s$-process}\label{sprocstars}

Neutron-capture element production through the $s$-process occurs in stars with 1 to 8\,M$_\odot$ during the last $\sim1\%$ of their lifetime. Stars with 1\,M$_\odot$ live for 10 billion years, whereas stars with 8\,M$_\odot$ only shine for just under 100 million years. Their last evolutionary stage is the so-called asymptotic giant branch (AGB) phase, during which such stars are thermally pulsating. In the AGB phase, a star has an inert carbon-oxygen core that is surrounded by the helium-burning shell, then a helium-rich inter-shell region, and a hydrogen-burning shell. Farther out is the convective envelope that reaches to the surface (\citep{herwig05} and references therein). These stars will ultimately die and become so-called planetary nebulae. Then, the outer gas layers are expelled to expose the hot core, a so-called white dwarf made from either carbon and oxygen, or oxygen, neon and magnesium. We now describe how the $s$-process operates in stars with specific stellar masses, as they produce elements with different atomic masses.

\textbf{The ``main'' $s$-process:} The dominant neutron source (leading to a neutron density of $n_{n}
\sim 10^{8}$ cm$^{-3}$) in low-mass stars with 1--3\,M$_\odot$
is the $^{13}$C($\alpha$,\textit{n})$^{16}$O reaction \citep{abia02}. However, this reaction can occur only if sufficient protons are mixed from the hydrogen-rich shell down into the helium inter-shell. Then, the so-called $^{13}$C pocket is activated, because the protons can combine with existing $^{12}$C to form $^{13}$C (through partial completion of the CN cycle) and produce neutrons. This series of processes eventually enables neutron-captures onto seed nuclei, such as Fe, to create $s$-process elements. Although many details relating to the existence and operation of the proton-mixing episodes and build up of the $^{13}$C pocket remain under debate (see \citep{herwig05} for a review), the $s$-process only operates with it in place \citep{gallino1998, goriely00,lugaro12, bisterzo17}. This process is called the ``main'' $s$-process because it provides a substantial neutron exposure (the time-integrated neutron density) which leads to the formation of elements with $Z>40$, in the second and third $s$-process peaks (see below), in between the convective thermal pulses, on timescales of 1,000 to 10,000 years. During each pulse, these newly created elements, as well as copious amounts of carbon, are dredged up to the stellar surface and eventually released into the surrounding interstellar gas through stellar winds. Given the large number of these stars (lower-mass stars are  more common than higher-mass stars), this stellar population contributes significantly to the the inventory of neutron-capture elements, as well as carbon, in the universe. 

\textbf{The ``weak'' $s$-process:} In stars with intermediate masses of 3 to 8\,M$_\odot$, the reaction $^{22}$Ne($\alpha$, n)$^{25}$Mg serves as the dominant neutron source \citep{abia02}. It follows $^{14}$N($\alpha$,$\gamma$)$^{18}$F($e^{+}+\nu$)$^{18}$O($\alpha$,$\gamma$)$^{22}$Ne, based on initial $^{14}$N stemming from the CNO cycle. It operates during the thermal pulses within the helium inter-shell on timescales of 10 years. The resulting neutron density is significantly higher than what is produced in low-mass stars, $n_{n} > 10^{10}$ cm$^{-3}$ \citep{herwig05, karakas12}, but the neutron exposure is reduced (due to the reduced timescale). The resulting $s$-process is called the ``weak'' $s$-process, and produces elements in the region of the first $s$-process peak, around $Z \sim 40$ \citep{busso_gallino_AGB1999}. This neutron source also operates during core helium- and shell carbon-burning in massive stars with $>12$\,M$\odot$ that eventually explode as supernovae. It produces a similar $s$-process-element distribution as the intermediate-mass AGB stars \citep{pignatari10}. Finally, it has also been suggested to operate in massive rotating ($\sim500$--$800$\,km\,s$^{-1}$ at the equator) near-metal-free stars \citep{Pignatari08,frischknecht12}. Fast rotation boosts $s$-process element production by several orders of magnitude (e.g., for Sr; \citealt{chiappini11} and references therein) because rotation-induced mixing increases the production of $^{14}$N, $^{13}$C and $^{22}$Ne, all of which aid in the operation of the $s$-process. The resulting yields provide qualitative explanations for unusual abundance patterns (e.g., $\mbox{[Sr/Fe]}= +1.2$ in HE~1327$-$2326 with $\mbox{[Fe/H]} = -5.4$; \citealt{HE1327_Nature}), and sub-solar [Sr/Ba] ratios in some metal-poor stars. However, more theoretical work may be needed to further refine predictions.

\begin{figure}
\centering
\begin{minipage}{.5\textwidth}
  \centering
  \includegraphics[width=2.9in]{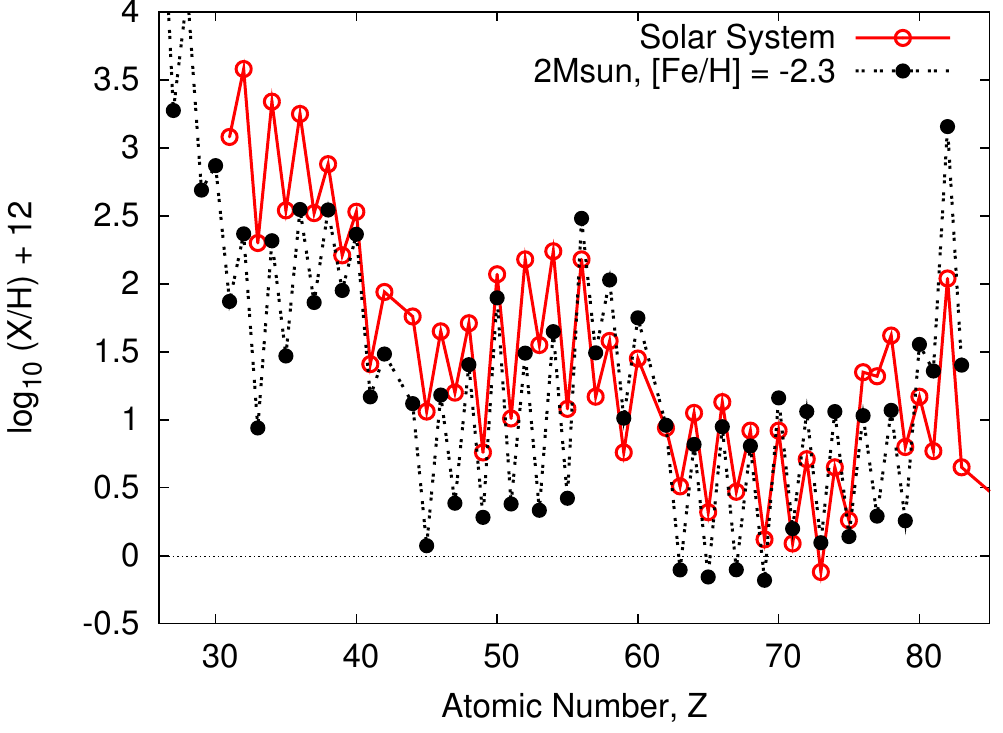}
\end{minipage}%
\begin{minipage}{.5\textwidth}
  \centering
  \includegraphics[width=2.8in]{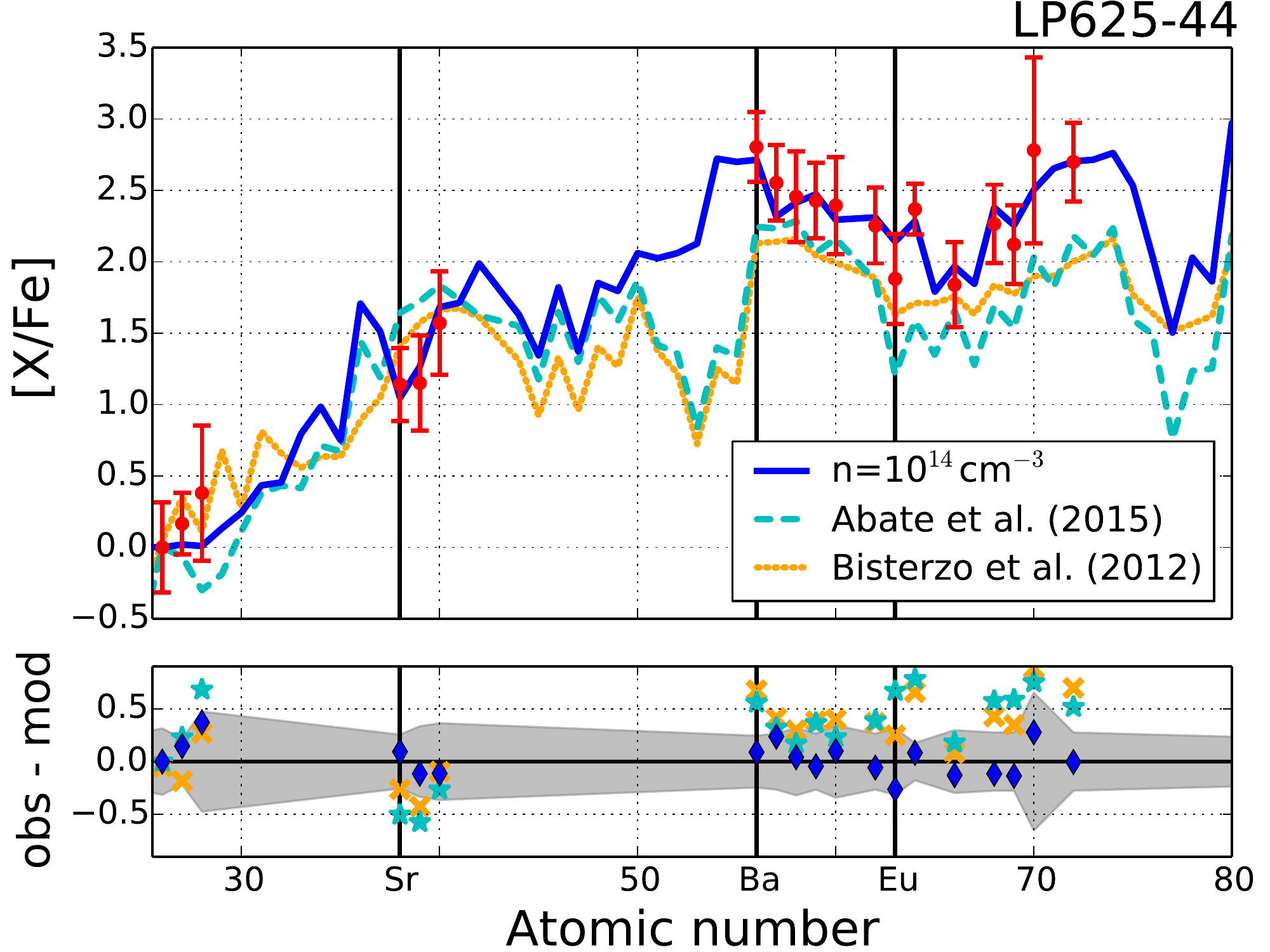}
\end{minipage}
\caption{\label{s_i_proc_fig} Left: The $s$-process (absolute) abundance
pattern inferred for the Sun compared with that calculated to be produced by a 2\,M$_\odot$ evolved star
with $\mbox{[Fe/H]} = -2.3$ \citep{lugaro12}. At low metallicity,
the $s$-process operates efficiently enough to produce $s$-process
abundances equal to or greater than the Solar System values,
including large amounts of lead ($Z = 82$). Right: Abundances (\textit{red circles}) of the star LP~625-44,
overlaid with nucleosynthesis models of the $s$-process (\textit{cyan})
\citep{Abate2015-2}, $s$-process plus an initial $r$-process
enhancement (\textit{orange}) \citep{Bisterzo12}, and the $i$-process (\textit{blue})
\citep{hampel16}. Residuals of models and data are shown at the
bottom. The gray area reflects observational uncertainties from
the top panel. Left panel  courtesy of A.
Karakas. Right panel  courtesy of M. Hampel.}
\end{figure}

The overall emergence of $s$-process material in star-forming gas is documented by the body of observations of metal-poor halo stars with metallicities of $\mbox{[Fe/H]} > -2.6$ that formed as part of the next stellar generation following the initiation of substantial $s$-process enrichment. We note that at lower metallicities, insufficient seed nuclei, such as Fe, are available for the $s$-process to operate, although earlier, individual contributions of $s$-process elements by short-lived massive stars is likely \citep{chiappini11}. Stars at various metallicities thus reflect consistently increasing enhancements in associated elements, such as barium, but of course also increasing amounts of other lighter elements, as part of ongoing chemical evolution. 

This highlights a central challenge for stellar archaeology. While these metal-poor stars clearly show the existence of $s$-process elements in their spectra, it is not possible for us to determine \textit{how many} progenitor stars contributed to the combined amount of $s$-process nuclei to the local gas prior to when these stars formed. A clean signature of the $s$-process thus cannot be obtained from ordinary metal-poor stars with $\mbox{[Fe/H]} > -2.6$. But fortunately, there is another way. Around 10\% of these metal-poor stars show extreme enhancements in $s$-process elements. An additional few such stars are also found in dwarf galaxies \citep{frebel14}. These low-mass stars happen to be part of binary star systems, with a companion star that is slightly more massive and thus more evolved (less-massive stars live longer than more-massive stars). The erstwhile companion is thought to have undergone the asymptotic giant branch phase, produced copious amounts of $s$-process elements (and C), and then transferred the enriched surface material to the now-observed star, where the transferred mass dominates the surface composition. Such mass-transfer events are common in binary systems of stars with orbital separations that are sufficiently small to experience Roche-Lobe overflow after one star (the more massive of the pair) inflates during late stellar evolution. This scenario has been confirmed by long-term monitoring of the radial velocities of such stars. Variations are found in 80\% of them \citep{hansen16_sproc_bin}, due to the observed star's motion around the (now usually unseen) companion. Naturally, large amounts of carbon are also transferred. What is then observed in these metal-poor stars is the characteristic signature of the main $s$-process element, together with a carbon enhancement, which are is the result of only \textit{one} progenitor, the companion star \citep{1997norriscarbon,bisterzo10}.

The signature of the main $s$-process can be easily recognized from the three tell-tale peaks in the abundance distribution, as function of atomic number $Z$ \citep{busso_gallino_AGB1999, sneden08}. The first peak is located at $Z=38-40$ (Sr, Y, Zr), the second peak at $Z=56-60$ (Ba through Nd), and the third peak at $Z=82-83$, which includes lead and bismuth, the termination point of the $s$-process, as shown in \textbf{Figure~\ref{s_i_proc_fig}}. Assuming sufficient neutron exposure, the peaks are the result of certain nuclei having low cross sections, and thus being particularly stable against neutron-capture. This happens for nuclei with a magic number of neutrons, namely 50, 82 and 126. Nuclei that have, in addition, a magic number of protons are the most stable, e.g. $^{208}$Pb with 126 neutrons and 82 protons. These nuclei become bottlenecks during $s$-process nucleosynthesis, and lead to the characteristic $s$-process peaks. At very low metallicity (e.g., $\mbox{[Fe/H]} \lesssim -2.0$), the observed nucleosynthesis pattern shows exactly this abundance distribution, including extreme lead abundances \citep{2001aokisprocess,ivans05,placco13}, as the $s$-process operates with a comparatively large neutron-to-seed ratio arising from the sparsity of, e.g., iron atoms. Then, the $s$-process runs to completion because there is no lack of neutrons, i.e., all the way up to lead. Predictions for the yields of low-metallicity $s$-process models are able to broadly reproduce the observed patterns, including the high lead \citep{karakas14}.

Somewhat different $s$-process patterns are produced in stars with progressively higher metallicity. This metallicity dependence of the $s$-process, as well as the fact that at least two different sites host $s$-process nucleosynthesis, implies that this process is not universal, but rather environment dependent. More metal-rich stars produce $s$-process signatures with much lower lead abundances. If cleanly observed, such as in a binary companion star, those patterns match the (slightly-scaled) solar pattern. We note that considering the total solar abundances at face value does not tell how much of each element (or each isotope) was made in the $s$-process, or $r$-process, over the course of 9 billion years prior to the Sun's formation. To obtain the solar $s$-process pattern, it has to be calculated. Given that $s$-process nucleosynthesis occurs along the valley of $\beta$ stability, neutron-capture rates and other nuclear properties have been successfully studied with experiments (contrary to the $r$-process that occurs far from stability, see Section~\ref{rprocstars}). Consequently, the $s$-process is theoretically well-understood \citep{busso_gallino_AGB1999, karakas11_models}, and the solar $s$-process component (or ``pattern'') has been calculated, also taking into account Galactic chemical evolution \citep{Burris00,Travaglio04, bisterzo17}.

\subsection{The astrophysical signature of the $i$-process}\label{iprocstars}

Some 20 CEMP stars are known with distinct
neutron-capture-element signatures that match neither the pattern
of the $s$-process nor that of the $r$-process. Therefore, they were
originally classified as CEMP-$r/s$ stars \citep{beers&christlieb05}. For
more than a decade, there has been extensively speculation as to whether a
combination of the two processes might explain the observations.
This group of objects then became known as the CEMP-$r+s$ or
CEMP-$s+r$ stars. However, no unambiguous explanation could be
found \citep{cohen2003, barbuy2005, jonsell06, cash1}. Instead,
given these stars' resemblance to $s$-process-enhanced stars with large
carbon abundances, recent theoretical research has begun to suggest
that a neutron source intermediate between the $s$- and $r$-processes
might be operating, and responsible for these observed chemical
signatures. This led to the (re)introduction of the so-called
intermediate neutron-capture process, or $i$-process \citep{dardelet15, hampel16}, following
an original suggestion by \citet{cowan_rose77}. This process
supposedly operates under neutron densities of $n_{n} \approx
10^{15}$\,cm$^{-3}$, several orders of magnitude higher than what
is required for the $s$-process ($n_{n} \sim 10^{8}$\,cm$^{-3}$; see
Section~\ref{sprocstars}), but likely varying neutron exposures
\citep{dardelet15}. Invoking a large neutron exposure
\citep{busso_gallino_AGB1999} results in an increased production
of heavy neutron-capture elements at and the second peak
(roughly $55<Z<75$), as is needed to match the observations.
Abundances of metal-poor stars with the CEMP-$r/s$ (or suggested
CEMP-$r+s$ or similar) label can be well reproduced with
nucleosynthesis calculations using neutron densities of $n_{n} \sim
10^{14}$\,cm$^{-3}$ \citep{hampel16}. By contrast, observations
of mainly first-peak elements in some post-AGB stars point to low
neutron exposure \citep{herwig11}. However, if the $i$-process is
truncated before reaching equilibrium, it can stop producing
elements at almost any atomic number, thereby producing few or even
no neutron-capture elements \citep{roederer16iproc}.

However, the astrophysical site(s) of this process remain in question. The best candidates are low- to intermediate-mass, low-metallicity AGB stars. They could potentially produce a significant burst of neutrons over sufficiently long timescales within the convective helium inter-shell region, if the necessary amounts of protons enter this region from above hydrogen-burning shell \citep{campbell08}. Alternatively, in the most massive ``super'' AGB stars \citep{doherty15}, the protons could enter during a core helium flash \citep{lugaro09,jones16}. Yet another possibility are rapidly accreting white dwarfs in close binary systems \citep{denissenkov17}. These protons could then combine with the available $^{12}$C to form $^{13}$C, so that the $^{13}$C($\alpha$,n)$^{16}$O reaction could produce a sufficiently high neutron density (up to $n_{n} = 10^{15}$\,cm$^{-3}$ \citep{cristallo15}) for the $i$-process to operate, similar to the $s$-process (Section~\ref{sprocstars}). The resulting nucleosynthesis yields would all arise from an environment with roughly similar neutron densities, and independent of the proton ingestion mechanism \citep{campbell10}. However, the neutron exposure would vary significantly with astrophysical site. The AGB stars would likely provide environments of high neutron exposures, whereas evolved massive stars would produce a low exposure during hydrostatic helium and carbon burning \citep{pignatari10}. 

Overall, there is ample observational data to suggest that the $i$-process has contributed to the composition (see \citep{roederer16iproc} for a full list), for example, in the post-AGB star ``Sakurai's object'' \citep{herwig11}, in pre-solar grains \citep{fujiya13}, and in metal-poor stars with super-solar [As/Ge] and solar or sub-solar [Se/As] ratios \citep{roederer16iproc}. More theoretical work is needed to firm up details on potential astrophysical sites, as well as yield predictions, to explain the full range of metal-poor stars with these unusual enhancements in neutron-capture elements.

\subsection{Star with signatures from both the $r$ and $s$-processes}\label{rs_stars}

\begin{figure}[!t]
\includegraphics[width=4.3in]{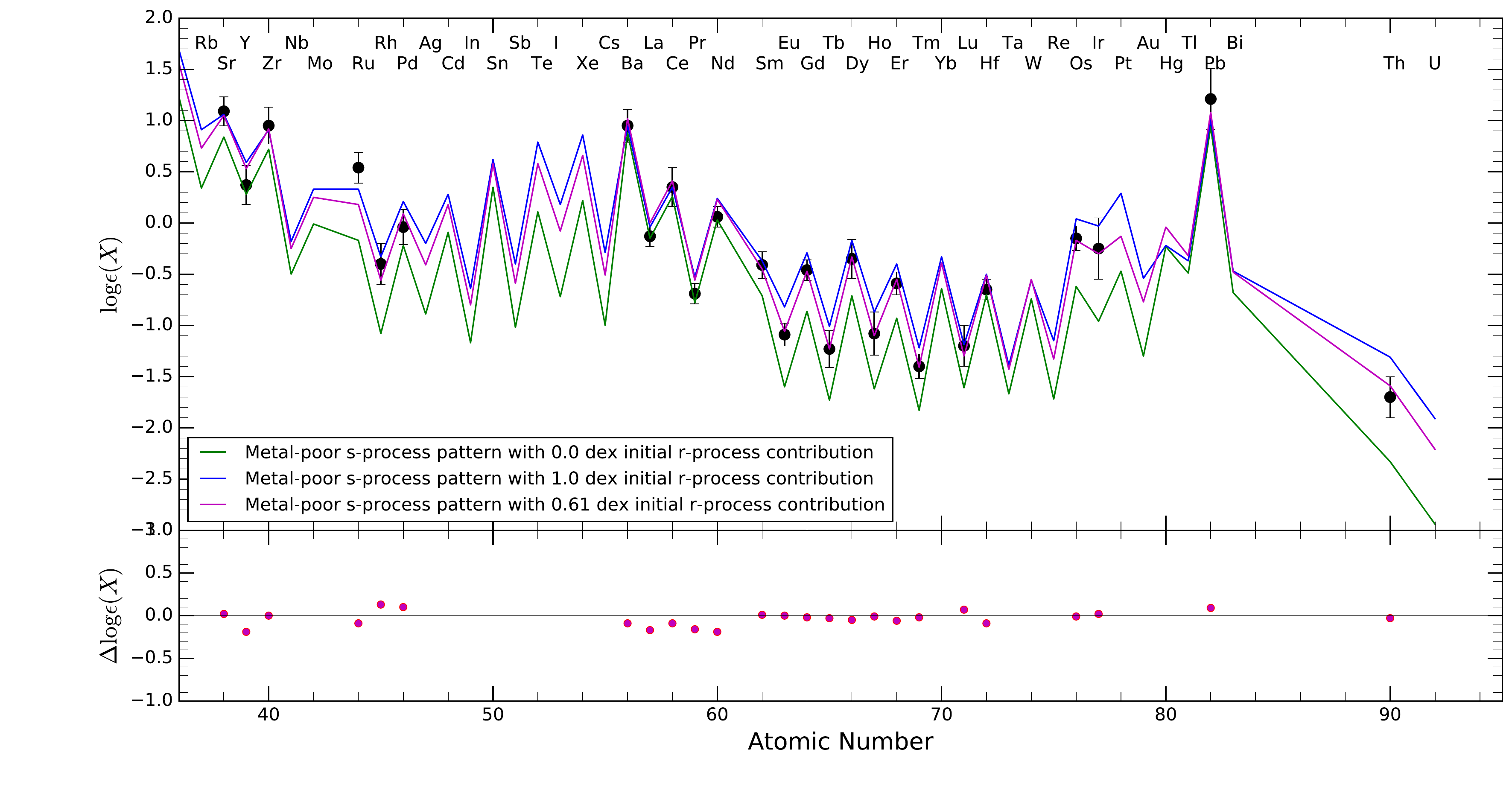}
\caption{\label{gull} Top: Abundances (black circles) of star RAVEJ0949$-$1617, overlaid with $s$-process yields to which different initial $r$-process enhancements have been added. Bottom: Residuals between the best-fit model and the observed data. Figure from \citet{gull18}.}
\end{figure}

In spite of the many attempts to explain the neutron-capture-element signatures of $i$-process-enhanced stars as a result of the {\it combination} of $s$- and $r$-process elements \citep{cohen2003, barbuy2005, jonsell06, cash1}, and the overall plausibility of this scenario, it has been surprising that no stars with this signature have been discovered. However, recent searches for $r$-process-enhanced stars \citep{hansen18, sakari18} have finally delivered a first example. \citet{gull18} describe a metal-poor giant star, RAVE~J0949$-$1617, with $\mbox{[Fe/H}]=-2.2$ and carbon enhancement ($\mbox{[C/Fe}]=1.2$), that displays a combination of enhanced $s$- and $r$-process elements. The likely scenario for stars such as RAVE~J0949$-$1617 is that it must have formed in a binary system from gas enriched by a prior $r$-process event. Later on, it received carbon and $s$-process material during a mass-transfer event from the binary companion star. A combination of isotopes made in the two processes is now observed in the form on an unusual neutron-capture-element pattern that is governed by high barium and lead abundances (due to the $s$-process component), and lower rare-Earth element values (from the $r$-process). \textbf{Figure~\ref{gull}} shows the abundance pattern of this star. Modeling the combination of the two processes, following \citet{Abate2015-2}, revealed an initial $r$-process enrichment of 0.6\,dex of the natal gas cloud. This makes the star a moderately $r$-process-enhanced star that happens to be in a binary system. We note that at least the signature of J0949$-$1617 is unlike that of the $i$-process when compared with model predictions \citep{hampel16}. While other combinations of the amounts of $r$- and $s$-process elements are possible, more of these stars need to be discovered to assess whether any similarities to $i$-process-enhanced stars exist.

For now, it remains puzzling as to why no other bona-fide CEMP-$r+s$ stars have yet been found. The fraction of low-mass metal-poor stars in binary systems is $\sim$16\% \citep{carney2003}; the fraction of moderately $r$-process-enhanced stars is also $\sim$15\% (see Section~\ref{rprocstars}). At face value, CEMP-$r+s$ stars should thus be rare, namely $\sim2$\%. But still, they should be about half as common as the rare highly $r$-process-enhanced stars, of which at least two dozen are known at present. So more than just one CEMP-$r+s$ should have been found by now. This could suggest that either the binary fraction was perhaps less than previously thought, that some process prevents the proper identification of CEMP-$r+s$ stars \citep{gull18}, or other reasons. While more of these stars are awaited from ongoing and future searches, for now, this one object defines the class of bona-fide 
CEMP-$r+s$ stars.  

\subsection{The Astrophysical Signature of the $r$-Process}\label{rprocstars}

Rapid neutron capture occurs when the neutron density reaches $n_{n} >
10^{22}$\,cm$^{-3}$. Then, seed nuclei are bombarded with neutrons
to form neutron-rich isotopes far from stability, including those
in the transuranian region, within only 1--2\,s. Once the
neutron flux stops, the nuclei decay to form stable isotopes,
including the actinide elements thorium ($Z=90$) and uranium
($Z=92$). The resulting abundance distribution is difficult to
predict ad hoc, but abundance results for the Sun have shown a
characteristic elemental pattern that is also found in metal-poor
stars with enhancements in neutron-capture elements
\citep{sneden08}. This pattern is also described by three abundance
peaks, but they are distinctly shifted to lower mass numbers,
compared with the $s$-process pattern. The first peak occurs at $Z =
34$--36 (selenium, bromine, krypton), the second at $Z = 52$--54 (tellurium, iodine, xenon), and
the third at $Z = 76$--78 (osmium, iridium, platinum). The peaks in the
$s$-process distribution are due to the pileup of stable nuclei with magic neutron numbers (see
Section~\ref{sprocstars}). In the $r$-process, the peaks result
from the decay of {unstable} neutron-rich nuclei with magic numbers of
neutrons, leading to peaks at lower mass numbers than the s-process.

Only a few astrophysical sites can produce such high neutron densities. Candidate sites are thus limited to core-collapse
supernovae, magnetorotationally jet-driven supernovae, and
mergers of a binary neutron star pair or a neutron star--black hole
pair (see \citealt{thielemann17} and references therein). Identifying this site (or sites) has remained a major challenge for nuclear astrophysics
since the 1950s, when researchers first recognized that the $r$-process is required to account for the solar abundances of heavy elements \citep{Burbidge57, cameron58}. Recent
astrophysics observations of old stars in a dwarf galaxy have
finally shown neutron star mergers to be the most likely
production site \citep{ji16b}. This example of the interplay of
astrophysics and nuclear physics to identify the site of the
$r$-process is discussed in more detail in
Section~\ref{sec4}. Here, I provide a general description of
the observational data on the $r$-process available through
metal-poor stars, and refer the reader to  \citet{sneden08} for a
discussion of the Sun's $r$-process pattern.

Ample evidence exists {that} the $r$-process operated in
the early Universe, specifically in the form of rare metal-poor stars
showing unusually large amounts of neutron-capture elements
associated with the $r$-process in their spectra. Their
corresponding elemental abundances exhibit the characteristic
signature of the $r$-process; examples are shown in
\textbf{Figure~\ref{comp}}. Dedicated searches for these
$r$-process-enhanced stars \citep{christliebetal04} found that they
are quite rare: 3--5\% of otherwise-ordinary metal-poor stars
with $\mbox{[Fe/H]}<-2.5$ in the Galactic halo contain a strong
enhancement of $r$-process elements \citep{heresII}. Recall
that these stars did not produce any of these elements themselves.
In fact, $r$-process enhancement is found across all major phases
of stellar evolution \citep{roederer14rproc_hbstars}, further
indicating that these elements are not the result of any peculiar
atmospheric chemistry. Based on long-term monitoring of their
radial velocities \citep{hansen15_rproc_bin}, the great majority of
$r$-process-enhanced metal-poor stars ($\sim$82\%) exhibit no
variations arising from the presence of a binary star companion.
Thus, mass transfer from the companion to the presently observed
star cannot be the cause of its $r$-process enhancement; rather,
 $r$-process-enhanced stars preserve the chemical
fingerprint of the nucleosynthesis processes that enriched their
birth gas clouds. These natal environments are expected to have
been enriched by core-collapse supernovae in light elements (found
in all metal-poor stars), as well as an $r$-process event that
provided the neutron-capture elements observed. The low
metallicity of the stars implies that $r$-process nucleosynthesis
must have already taken place in the early Universe, in
environments with limited additional star formation, so as to not
erase the distinctive $r$-process patterns. Only one (or very few)
progenitor supernova would be needed to produce the (small) amount
of iron found in extremely metal-poor stars with $\mbox{[Fe/H]}
\sim -3.0$. 

As for the $r$-process elements, one can assume, in a
first instance, that also just one progenitor event produced them.
This makes these stars particularly interesting for stellar
archaeology. However, stars that challenge this notion (those with
the actinide boost; see below) are also being found
\citep{Hill02,ji18}. These cases might be due to
the $r$-process operating in multiple sites, or variations of
nucleosynthetic pathways within a given site. The question of
progenitor objects and events is further complicated by
suggestions that there may have been differences in the
birth environments and the associated levels of $r$-process yield
dilution within the natal gas clouds from which these
$r$-process-enhanced stars formed. This idea is supported by
$r$-process-enhanced stars with higher [Fe/H] values. They may
simply have formed in larger systems in which the $r$-process was
diluted more, or from gas enriched by multiple progenitors objects
and/or events.

The europium abundance relative to iron, [Eu/Fe], is commonly used to quantify the
level of $r$-process enhancement ({Table~\ref{categories}}).
Two levels of $r$-process enrichment are (arbitrarily)
distinguished for convenience. The strongly enhanced $r$-II stars
have $\mbox{[Eu/Fe]} > +1.0$, whereas the moderately
$r$-process-enhanced $r$-I stars show $+0.3 \le \mbox{[Eu/Fe]} \le
+1.0$. The group of $r$-II stars thus exhibits europium-to-iron ratios more than 10 times that of the Sun, indicating that, compared with
lighter elements such as iron (which determine the metallicity of
the star), unusual amounts of $r$-process elements are present.
Approximately 30 of them are currently known. Many more $r$-I stars,
around 130, have been discovered to date, corresponding to
$\sim$15\% of metal-poor stars. Examples of well-studied $r$-II
stars include the very first one to have been discovered,
CS~22892-052 \citep{sneden2000}, which is also a carbon-enhanced
star; the first one with a uranium measurement, CS~31082-001
\citep{Cayreletal:2001}, which also exhibits an actinide boost (see Section 3.4.4); and HE~1523$-$0901, which also has a uranium measurement
\citep{frebel07}. All three stars have $\mbox{[Fe/H]} \sim -3.0$,
but the neutron-capture material is 40--70 times more abundant
relative to iron. Interestingly, the metallicity distribution of
both $r$-I and $r$-II stars broadly covers the range of $-3.5 <
\mbox{[Fe/H]} \lesssim -1.5$. However, at $\mbox{[Fe/H]} \sim
-1.5$, the median [Eu/Fe] of halo stars is
$\mbox{[Eu/Fe]} \sim +0.3$. This is where the classification of
$r$-process-enhanced stars ceases to be useful, since nearly
$50$\% of halo stars have $\mbox{[Eu/Fe]} > 0.3$ and would thus be
$r$-I stars. In these cases, only stars showing the $r$-process
pattern can be identified as truly $r$-process-enhanced stars. In
this context, note that metal-poor stars with
$\mbox{[Fe/H]}<-2.6$ naturally show neutron-capture-element
signatures free of any $s$-process contribution, since the
$s$-process begins to contribute significantly to the
chemical enrichment of star-forming gas only around
$\mbox{[Fe/H]}\sim-2.6$ \citep{simmerer04}. At
$\mbox{[Fe/H]}>-1.4$, the $r$-process stops dominating the
chemical evolution of neutron-capture elements \citep{roederer10b}.

I now describe three different groups of astrophysical conditions and sites that enable $r$-process nucleosynthesis (details are given in {Table~2)}. This description of the underlying physics processes will help  us to understand different portions of the observed $r$-process
abundance pattern that may arise from contributions produced
by different astrophysical sites.

\subsubsection{The early phase of any $r$-process: QSE, and hot vs. cold $r$-process}

One form of the $r$-process, the so-called hot $r$-process, undergoes an initial quasi-statistical equilibrium (QSE) phase, which typically creates elements from strontium to silver, before significant neutron capture takes place (the case of a stand-alone terminal QSE is also possible under conditions that prevent any additional neutron capture to create elements heavier than silver). A hot $r$-process operates at $T > 10^{9}$\,K, when hot and dense material initially consisting of protons and neutrons is expanding, such as in any kinds of wind emerging from a supernova or neutron star merger environment. During the QSE phase, seed nuclei are created close to stability, from $\alpha$, neutron, and proton captures and the reverse processes. Neutron capture eventually occurs onto these newly created seeds, such as strontium, during the subsequent $r$-process, but conditions are not suitably neutron rich to consistently produce elements heavier than silver.  

In contrast, the cold $r$-process operates at $T\sim 10^{8}$\,K. Neutrons become available, e.g., when ($\alpha$,$n$) reactions are activated by shock heating in a supernova. Subsequent neutron capture occurs onto preexisting seeds, such as iron, in the birth material of the supernova progenitor. Due to the lower neutron density, the neutron-rich nuclei produced this way are much closer to the valley of $\beta$ stability compared with nuclei made in the hot $r$-process.

\subsubsection{The main $r$-process}

\begin{figure}[!t]
\includegraphics[width=4.6in,clip=]{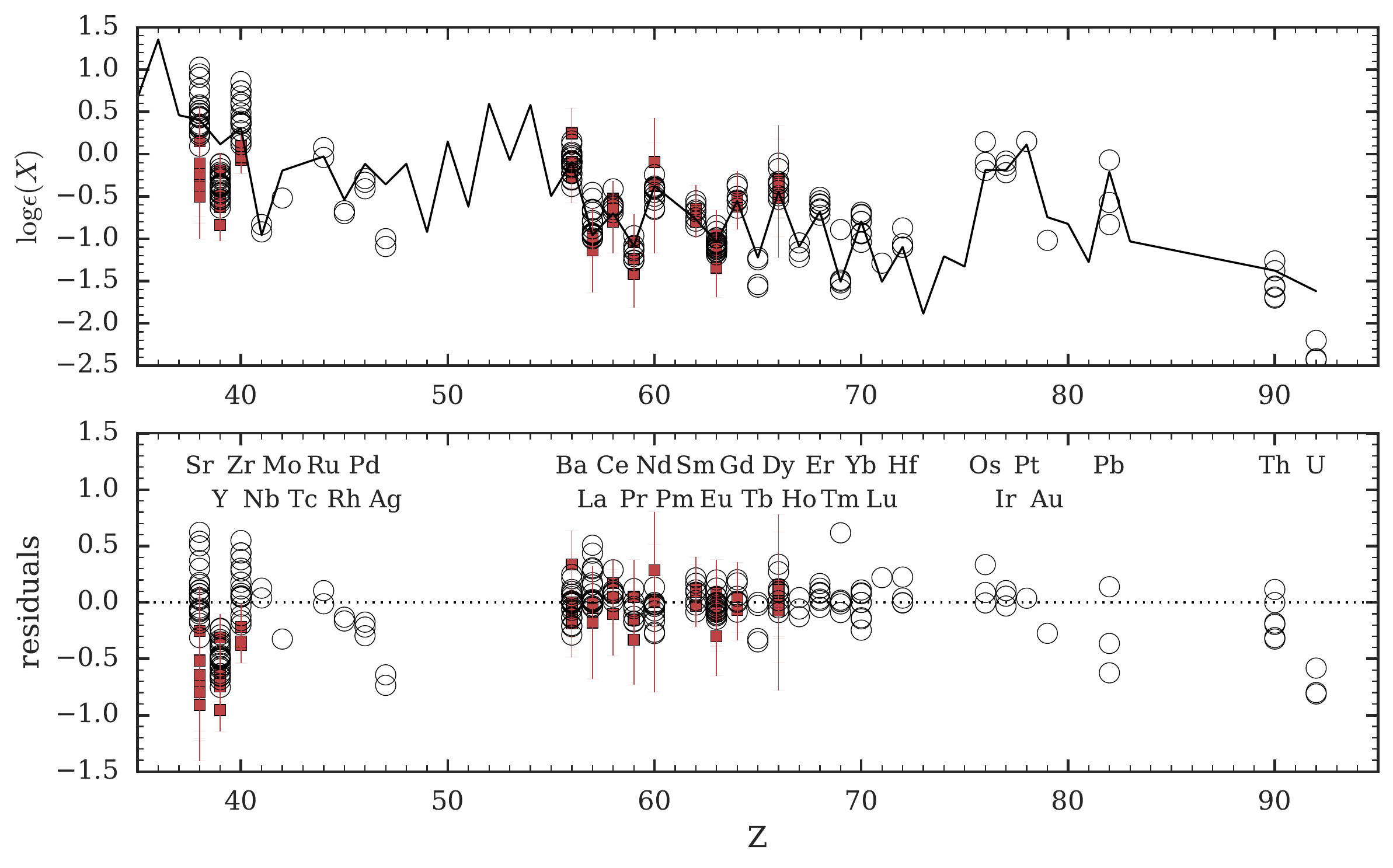}
\caption{\label{pattern_plot} (\textit{a}) Normalized $r$-process-element
abundances of metal-poor halo (\textit{open circles}) and Reticulum\,II
stars (\textit{red squares}), overlaid with the scaled solar $r$-process
pattern (\textit{line}) \citep{Burris00}. (\textit{b}) Differences between the
stellar abundances and the solar pattern. Figure courtesy of A.
Ji.}
\end{figure}

It has been well established that the observed abundances of
$r$-process elements in $r$-process-enhanced stars display
essentially the same relative pattern for elements barium and
above, even though the absolute
enhancement levels of these elements vary by $\sim$1.5 orders of
magnitude (as evidenced by, e.g., the range of europium and the associated iron abundances). \textbf{Figure~\ref{pattern_plot}} displays rare-earth
elements with $56 \le Z \le 70$ in various $r$-process-enhanced
stars. These abundance patterns show little scatter, in contrast to the
lighter elements around $Z\sim40$. Remarkably, the stellar
$r$-process patterns are also nearly identical to the scaled solar
$r$-process component \citep{Burris00} that can be extracted from
the total solar abundances by subtracting the
theoretically calculated $s$-process component (see
Section~\ref{sprocstars}). Given that the Sun formed billions of
years after these metal-poor stars, from gas that was enriched by
many stellar generations in various ways, the astounding
agreement between the patterns suggests that the $r$-process is
universal, at least for elements in and above the second peak
\citep{roederer12b}. No matter where and when the $r$-process
occurs, the elemental signature appears to be robust. This is
referred to as the main $r$-process. Elements with $Z>70$ show
somewhat more scatter, which is at least partially due to
observational uncertainties. Thorium and uranium, of course, canonically differ from the scaled solar pattern due to their decay (with actinide deficient/actinide
boost stars as exceptions; see Section 3.4.4).

\begin{table}[!t]
\caption{\label{Tab:rprocs} Nucleosynthesis processes that can
contribute neutron-capture elements}
\begin{tabular}{|l|c|c|c|c|}
\hline
Process & Conditions & Elements & Y$_{e}$& Astrophysical sites \\
&& produced&& \\
\hline
Terminal & Insufficiently neutron rich; & Sr $\rightarrow$ Ag&$<0.5$ & Standard proto--neutron \\
QSE$^{\rm{a}}$ & $\alpha$-, neutron-, proton-capture && & star wind in core-collapse\\
& and reverse; expansion from && & supernovae;\\
& hot, dense state & & &shock-heated/disk ejecta \\
\hline
$\nu$\textit{p}-process & Proton rich, $\xoverline{\nu}_{e}$ rich; & Sr $\rightarrow$ Ag &$>0.5$ & Standard proto--neutron\\
& QSE and $\xoverline{\nu}_{e}$ capture &  & & star wind in core-collapse supernovae;\\
& & & &shock-heated/disk ejecta \\
\hline
Limited$^{\rm{b}}$ & Neutron-to-seed ratio $<<100;$ & Sr $\rightarrow$ Ba&$<0.5$& Modified proto--neutron star wind; neutron \\
$r$-process & QSE and & (limited && star merger: disk (after merger, viscous/\\
 & (limited) neutron capture; & production) && wind timescales); shock-heated ejecta\\
 & no fission cycling &toward Ba & & (during merger, dynamical timescales) \\
\hline
Main & $\mbox{Neutron-to-seed ratio}>100;$ & Ba$\rightarrow$ U &$<0.2$& Neutron star merger: tidal ejecta \\
$r$-process & QSE and neutron capture; & & & (during interaction);\\
& any fission cycling & & &dynamical ejecta (during merger)\\
\hline
Robust & $\mbox{Neutron-to-seed ratio}>100$; & Ba $\rightarrow$ U&$<0.2$& Neutron star merger: tidal ejecta \\
(main) &QSE and neutron capture; &&&(during interaction); \\
$r$-process &fission cycling limit & & & dynamical ejecta (during merger)\\
\hline
\end{tabular}
\begin{tabnote}
$^{\rm{a}}$Quasi-statistical equilibrium; see  \citet{meyer98} for a detailed description and treatment.\\
$^{\rm{b}}$Often referred to as the weak $r$-process or the light-element
primary process (LEPP). However, the term ``weak'' does not well describe
the nature of the underlying $r$-process physics, and ``LEPP''
does not refer to a specific nuclear physics process.
\ 
\end{tabnote}
\end{table}

Note that even though measurements of isotope abundances would
be very insightful for studying the universality of the main
$r$-process, they are impossible to determine for most elements,
for various reasons \citep{roederer08iso}. The only exceptions
are barium, europium, and possibly neodymium and samarium, but
results are often too uncertain to actually provide stringent
constraints on their (isotopic) formation process
\citep{gallagher12}. In addition, it is astounding to recall that
the $r$-pattern was discovered on the basis of the derivation of 1D LTE
abundances, despite
the range of metallicities covered by the stars. The stability and reproducibility of the pattern
imply that systematic abundance uncertainties, such as NLTE or 3D
effects, cannot be of a significant differential nature for ionized species in the rare-earth regime, although these
heavy elements might still be equally affected (which would simply
shift uniformly but not differentially change the pattern). The universality of the main
$r$-process offers a unique opportunity to provide
observational constraints on theoretical modeling of the
$r$-process because the stars clearly suggest only one end
result. This enormous advantage makes $r$-process-enhanced stars
ideal test objects for nuclear physics, complementing nuclear physics
experiments, which cannot yet reach the heaviest neutron-rich nuclei involved in the $r$-process.

But which site can produce this end result? Generally, a very neutron rich environment with an electron fraction of $Y_e < 0.2$ is required to produce the second- and third-peak $r$-process elements. The electron fraction, $Y_e =1/(1+N_n/N_p)$, is the ratio of
electrons to baryons (i.e., neutrons and protons) describes
the neutron richness, and is a critical parameter that ultimately determines which elements are made \citep{Metzger10,goriely11}. It changes when protons capture electrons to form neutrons and, thus, with the environmental conditions where an $r$-process can occur. Accordingly, moderately neutron- or proton-rich QSE conditions with $Y_e \gtrsim 0.5$ (when $N_n/N_p<1$) 
enable the synthesis of first-peak elements. The main $r$-process thus operates under $Y_e < 0.2$ conditions that become as extreme as $Y_e \sim 0.05$. In such a neutron-rich environment, with a neutron-to-seed ratio of $>>100$, fission cycling
occurs before the $r$-process freezes out. This fission cycling nuclei in
the second-peak region are produced from the fission of neutron-rich nuclei in
the transuranian region. The fission
products then become seed nuclei themselves, contributing again to the
formation of elements barium to uranium. Depending on the amount
of fission cycling, the main $r$-process can be split into two
components. A standard main $r$-process feeds on any amount of
fission cycling, whereas a particularly robust main $r$-process
runs at the fission cycling limit, i.e. a constant fission-induced production of second-peak elements. Both of these $r$-processes create elements barium to
uranium, and produce the kind of universal $r$-process patterns
that are found among $r$-process-enhanced metal-poor stars and the
Sun. {Table~2} lists the various nucleosynthesis processes and
conditions that produce certain element groups as part of what is
generically called the $r$-process. Astrophysical sites where an
$r$-process can operate are described below.

The astrophysical site now believed to be the prime candidate for
hosting any type of main $r$-process is during the merger of a
pair of orbiting neutron stars. Neutron stars are the compact
remnants of stars with masses of 10--20\,M$_\odot$. High-mass
stars were likely dominant early on in the Universe, with a
propensity to be born as binaries or multiples \citep{demink15}.
Modeling of a main $r$-process occurring during the very complex
merger process (consisting of different phases and components) has
been extensive (see \citealt{fernandez16,thielemann17,rosswog99,wanajo14,Lippuner15} and references therein) because conditions are
generally very favorable, especially in the initially (cold) dynamical
ejecta that occur during the merger phase \citep{Eichler15}. Since
these ejecta all have $Y_e < 0.1$, only second- and third-peak
elements with $Z > 50$ are made \citep{Korobkin12}. The
corresponding predicted high elemental yield of these ejecta is
principally commensurate with the overabundances of heavy
$r$-process elements found in $r$-II stars \citep{ji16b}.

Additional $r$-process production occurs in the neutrino-rich
winds emanating from the disk surrounding the merged object. These
disk winds develop through a combination of viscous heating and
nuclear heating following $\alpha$-particle formation, and might
be able to eject more material than the dynamical ejecta
\citep{Wu16}. Weak interactions might significantly increase
the $Y_e$ of the disk material, which could result in the
production of all three peaks of the $r$-process elements, but
many parameters influence the eventual $Y_e$
distribution. The observed nucleosynthetic signature of a neutron
star merger is, in all likelihood, a combination of these two
types of ejecta, with disk wind yields being less robust than
those of the dynamical ejecta. The
resulting, observable $r$-process signatures may show at least
some scatter, in particular among first-peak elements. Halo star
observations might thus be able to place stringent constraints on
the $Y_e$ distribution of the progenitor $r$-process event and
possibly even its components. This topic is discussed further below and
in Section~\ref{sec4}. For completeness, I note that a similar
heavy-element pattern was predicted to be produced by
magnetorotationally driven jet supernovae \citep{Winteler12}.
However, new calculations cast doubt on the feasibility of this
scenario \citep{moesta17} and on the previously obtained high
$r$-process yield (similar to that of a neutron star merger), suggesting that these types of supernovae are not likely to be the main
producers of $r$-process elements in the early Universe.

\subsubsection{The limited $r$-process}\label{weakr}

While the $r$-process pattern appears to be universal for elements
in and between the second and third peaks, deviations among
first-peak elements and actinide elements from, for  instance, the solar
pattern are regularly observed in $r$-process-enhanced stars.
Specifically, a few $r$-process enhanced metal-poor stars display
first-peak abundances higher than the solar pattern (when scaled
to heavier elements such as europium), whereas most stars have lower
values. \textbf{Figure~\ref{pattern_plot}} illustrates these relative
first-peak deviations. In the case of strontium, they cover a
significant 1.5\,dex spread around the solar pattern. Reasons for
these discrepancies remain unknown, but the observed spread has
been interpreted as evidence for multiple $r$-processes or
$r$-process sites \citep{sneden2000,Travaglio04, Hill02, Kratz07,
Hill17}. A kind of ``failed'' $r$-process would be able to produce
only light neutron-capture elements with $Z<56$, which led to the
terms ``weak $r$-process'' \citep{wanajo01} and ``light-element
primary process'' (LEPP; \citealt{Travaglio04}). Instead, by
focusing on the underlying conditions of the $r$-process,
 this type of nucleosynthesis should be referred to as the
limited $r$-process due to a limited neutron-capture rate, as shown in {Table~2}. 

From a theoretical
point of view, the limited $r$-process begins with a phase of QSE. An
insufficient neutron-to-seed ratio of $<<100$ then enables only limited neutron capture to take place,
thus preventing the production of the heaviest elements.
Consequently, only light
neutron-capture elements around the first peak are made, with a rapidly
decreasing production toward second-peak elements. Generally,
this process might operate in (modified) neutrino-driven winds
emerging from a proto--neutron star formed after a core-collapse
supernova in a progenitor star of perhaps 8--10\,M$_{\odot}$
\citep{qian_wasserburg03,wanajo01}. Alternative sites are winds from the
accretion disks that form after the merger of two neutron stars,
or the shock-heated ejecta during such a merger. For
completeness, I note here that the $\nu$\textit{p}-process yields an element distribution very similar to that of the limited $r$-process. It
plausibly operates in more standard proto--neutron star neutrino-driven
winds, as well as during neutron star mergers, similar to the
limited $r$-process. But overall, the initial conditions are
fundamentally different for the $\nu$\textit{p}-process, as it requires a proton- and $\xoverline{\nu}_{e}$-rich environment with $Y_e>0.5$.

Having the limited $r$-process operate in core-collapse supernovae
has a number of advantages. If the main $r$-process makes second-
and third-peak elements, then the first-peak element variations found
in $r$-process-enhanced stars could be understood in terms of
supernovae providing variable amounts of light neutron-capture
elements to the gas from which these stars formed. Consequently,
the observed patterns of $r$-process-enhanced stars must be a
superposition of the yields of the limited {and} the main
$r$-processes. In addition, the supernovae would also be
responsible for the lighter fusion elements observed in
$r$-process-enhanced metal-poor stars. Also, since all other ordinary
metal-poor stars contain at least small amounts of neutron-capture
elements \citep{Roederer13}, the associated huge variations at a
given metallicity could be better understood if supernovae provide
this material as part of an ongoing chemical evolution. An important 
question is whether an isolated, clean signature of the limited
$r$-process might be observable, beyond the hints of its existence
provided by $r$-process-enhanced stars and their first-peak
deviations. The best candidate is the neutron-capture-element-poor
star HD122563 \citep{honda06}, which exhibits a relative
overabundance of light, first-peak neutron-capture elements
compared with heavier ones (where the overall level of enrichment is
rather low, commensurate with the low neutron-capture abundances
found in metal-poor stars). However, while rapid neutron capture
provides a plausible partial explanation, a significant contribution to
elements with $Z \gtrsim 38$ at early times is also predicted to
come from massive ($>8\,M_{\odot}$) stars, by charged-particle reactions associated with QSE in core collapse supernovae and/or by the
$s$-process \citep{pignatari10,chiappini11}.

Additional theoretical, experimental, and observational
studies are needed to better understand the details of
different $r$-process contributions, their sites, and their impact
on chemical evolution, including understanding how $r$-process
calculations are affected by uncertainties in the nuclear
properties (e.g., masses, neutron-capture cross sections,
$\beta$ decay rates) and the astrophysical conditions that enable
nucleosynthesis (e.g., $Y_{e}$, temperature, density, and expansion timescales; \citealt{Eichler15}). In
particular, uncertainties in the masses of neutron-rich nuclei
influence reaction rates, the $r$-process path, and the fission
region (and thus second-peak properties and other freeze-out
effects). Regardless, it is likely that multiple or even all of the 
$r$-process components listed in {Table~2} are ultimately required
to fully explain the complete neutron-capture-element abundance
patterns in $r$-process-enhanced stars, including the deviations
of the light neutron-capture elements with respect to the scaled solar $r$-process pattern.

\subsubsection{Yet another site? The actinide boost stars}

Some 30\% of $r$-II stars exhibit unusually high thorium (and
uranium) abundances \citep{heres5}, compared with the other stable
$r$-process elements such as europium, leading to increased abundance
ratios that are three to four times higher than those of other
$r$-process-enhanced stars. Interestingly, since the actinide
elements appear to be equally affected \citep{roederer09} by this
phenomenon, the ratio of uranium to thorium still provides a reasonable age of 14 billion years for CS~31082-001, even though its ratio
of thorium to europium suggests a {negative} age (see Section~\ref{ages}). This behavior has been termed an actinide boost
\citep{schatz_chronometers}. Its origin remains unknown, but
speculations include multiple $r$-process sites with different
conditions \citep{Hill02, Kratz07}. Interestingly, the brightest star in Reticulum\,II shows actinide \textit{deficiency} which further adds to this discussion (see Section~4.1).
 Regardless of whichever
conditions and/or astrophysical $r$-process site(s) are operating,
the frequent occurrence of the actinide boost stars will somehow
need to be explained.


\subsubsection{Nucleo-Chronometric age dating of the oldest stars}\label{ages} 

The heaviest long-lived elements made in the $r$-process are the radioactive isotopes $^{232}$Th and $^{238}$U. Thorium has a half-life of $14$ billion years; uranium of $4.5$ billion years. An abundance for thorium, as well as those of other stable $r$-process elements including europium, can be readily measured in the spectra of $r$-process-enhanced stars. Uranium, on the other hand, is very challenging to detect, given the weakness of the one available optical line, which is also severely blended by other absorption features. To date the event that created these isotopes, the abundance ratio of a radioactive element (i.e., Th, U) to a stable $r$-process element (e.g., Eu, Os, Ir) is compared to the respective initial $r$-process production ratio \citep{schatz_chronometers,Hill17}, analogous to dating archaeological artifacts through radiocarbon analysis. Ages can be calculated with the following equations \citep{Cayreletal:2001}:

 $\Delta t = 46.78\cdot[\log (\rm{Th/r})_{\rm initial} - \log \epsilon(Th/r)_{now}]$\\ 
\indent $\Delta t = 14.84\cdot[\log (\rm{U/r})_{\rm initial} - \log \epsilon(Th/r)_{now}]$ \\
\indent $\Delta t = 21.76\cdot[\log (\rm{U/Th})_{\rm initial} - \log \epsilon(Th/r)_{now}]$ \\ 

\begin{figure}[!t]
\begin{center}
\includegraphics[width=5in,clip=]{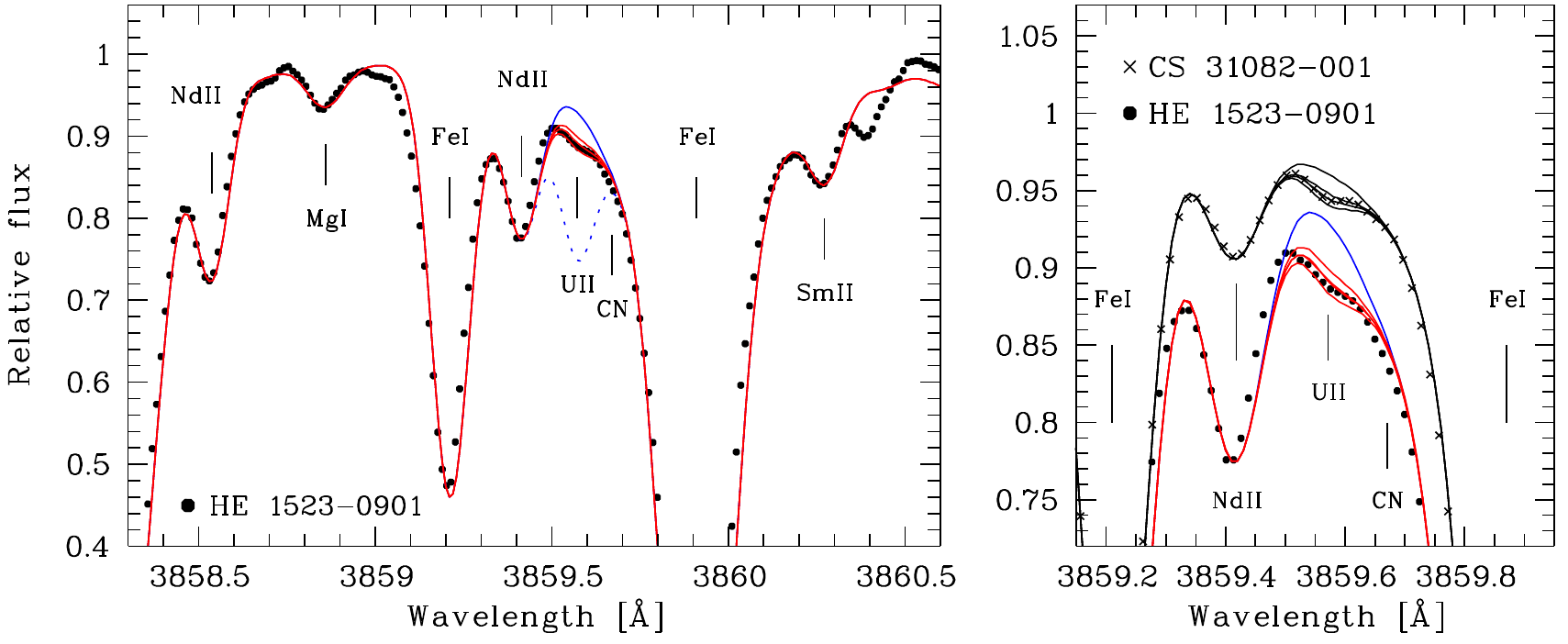} 
  \caption{Spectral region around the U\,II line at 385.9\,nm in HE~1523$-$0901 (black dots) and CS~31082-001 (crosses; right panel). Overplotted are synthetic
  spectra with different U abundances. The blue dotted line indicates the line strength if uranium had not decayed. Figure  from \citet{he1523}.}\label{U_region} 
\end{center}
\end{figure}

In this manner, about 20 $r$-process-enhanced stars with thorium detections have been dated to $\Delta t$ = 10-14 billion years, depending on which abundance ratio was employed \citep{johnson_bolte,sneden2000, cowan_U_02,heresII,hayek09}. Regarding uranium, only five stars have reliable abundance measurements \citep{Cayreletal:2001,frebel07,Hill17,Placco17}. 
\textbf{Figure~\ref{U_region}} shows the spectral region around the weak uranium line in HE~1523$-$0901, which is currently the most reliable uranium measurement. Ideally, abundances of both radioactive elements are available to produce a U/Th ratio, as well as many ratios with the stable elements so that ages from all the different ratios can be calculated. As can be seen from the equations, the Th/r ratio is the least accurate. The U/Th ratio should be less susceptible to uncertainties in the nuclear physics inputs required to calculating the production ratio. Uncertainties are expected to largely cancel out because the two elements have nearly the same atomic mass \citep{wanajo2002, kratz2004}. For the same reason, the heaviest stable elements from the third $r$-process peak region (Os, Ir) are more desirable to use for the age dating than lighter ones, such as europium. This is somewhat unfortunate, since europium is the most easily measured elements of the ones typically used for age dating. Accordingly, stars with both thorium and uranium are the most rare but also most-desired $r$-process-enhanced stars. Unfortunately, realistic age uncertainties range from $\sim2$ to $\sim5$\,Gyr depending on which ratio is used (see \citealt{schatz_chronometers} and \citealt{he1523}) for more detailed  discussions). Another general issue with existing production ratios is that they are based on $r$-process calculations tailored to supernovae being the astrophysical site of the $r$-process, or are simply site-independent calculations. Neither might be a good match, since neutron star mergers are likely the primary production site (Section~4), but also because production ratios might be individual to each event. Multiple $r$-process-enhanced stars indicate this \citep{Hill17, ji18}; more theoretical work is clearly required to obtain accurate and precise stellar ages from this approach.

The good news is that the many different $r$-process models, all
based on various assumptions out of necessity, can principally
be self-consistently constrained by $r$-process-enhanced stars if
abundances of all three of the elements thorium, uranium, and lead are
available. These elements are intimately coupled not only to
one another but to the conditions under which they were formed.
Lead is the $\beta$ plus $\alpha$ decay end product of all decay
chains in the mass region between lead and the onset of dominant
spontaneous fission in the transuranian region. In addition, a
portion of it is produced by the decay of uranium isotopes (and,
less so, by thorium). The isotope $^{238}{\rm U}$ decays into
$^{206}{\rm Pb}$, $^{232}{\rm Th}$ into $^{208}{\rm Pb}$, and
$^{235}{\rm U}$ into $^{207}{\rm Pb}$ (where the last one is based
on a theoretically derived ratio of $^{235}{\rm U}/^{238}{\rm
U}$). Thus, known abundances of all three elements provide the
only available observational constraint on the still
poorly understood actinide production. Models of the
$r$-process will need to be able to successfully reproduce those three
abundances. 

These same models might also be used to provide
improved initial production ratios that could then increase the
accuracy of stellar age dating. Currently, two $r$-II stars
have lead measurements available, CS~31082-001 \citep{plez04} and
HE~1523$-$0901 (A. Frebel et al., in preparation) only a few of the
$r$-process-enhanced stars are suitable for abundance measurements
of thorium, uranium, {and} lead. These elements have their
strongest optical lines at 401.9\,nm, 385.9\,nm, and 405.7\,nm,
respectively. Overall, the cooler the star is, the stronger the
absorption lines appear in the spectrum (at fixed abundance).
Of course, increasing the elemental abundance also increases the
line strengths, which is generally desirable when attempting to
measure elements that have only weak atomic transitions. Many
neutron-capture-element lines are in fact very weak and blended,
and thus require exceptionally high quality spectra ($R>60,000$
with signal-to-noise of $S/N>350$ per pixel at 390\,nm for uranium measurement, and
$S/N\sim500$ at 400\,nm for lead). In addition, the carbon
abundances should be low (ideally, subsolar) to minimize blending
of all three lines with CH and CN features; otherwise, measurements
are not possible. Finally, bright (preferably $V < 13$ magnitudes) metal-poor
giants are most desirable in this regard, in order to collect the
required spectra in reasonable observing time. As more bright
$r$-process-enhanced stars are discovered \citep{hansen18,sakari18}, at least a few more lead measurements can hopefully be attempted soon.
 
Regardless of all these challenges, the old ages derived for $r$-process-enhanced stars qualitatively confirm that low-metallicity stars are indeed ancient, and formed shortly after the Big Bang. In the absence of a reliable age-metallicity relation for halo stars, this implies that other metal-poor stars without any $r$-process enhancement are also old, as there is no discernible difference in the light-element abundances of $r$-process-enhanced stars and ordinary metal-poor stars. After all, $r$-process-element production is independent from that of the light fusion elements. The commonly made assumption about the low mass (0.6 to 0.8\,$M_{\odot}$) implying a star's current old age also appears justified, and the concept of stellar archaeology, namely that metal-poor stars are suitable for studying the conditions of the early universe, is broadly validated. In addition, the 10-14 billion-year-old $r$-process-enhanced stars provide a lower limit on the age of the Milky Way and the Universe itself. An age of the Universe of 13.8\,Gyr has been derived from observations of the cosmic microwave background radiation, as interpreted with the latest cosmological models \citep{PlanckCosmology}. Prior to the discovery of $r$-process-enhanced stars and precision cosmology, globular clusters were thought to be the oldest observable objects, with ages around 12 to 16 billion years. This was roughly in line with the age of the universe known at the time.

\section{DWARF GALAXIES AS NUCLEAR PHYSICS LABORATORIES}\label{sec4}

A cornerstone of nuclear astrophysics is establishing the astrophysical sources and sites of 
heavy-element production. The work with $r$-process-enhanced metal-poor stars has provided much insight into \textit{that} the $r$-process occurred already in the early universe, likely only sporadically, given the small fraction of $r$-process-enhanced stars. We also learned that it produces a universal pattern among the heavy neutron-capture elements and that challenges exist in the first-peak region and in relation to the actinides. The next step includes employing astrophysical evidence to gain insight into \textit{where} this process occurs, as the halo stars, with their (individually) unknown origins, cannot provide this information. In contrast to the sites of all other major nucleosynthesis processes being relatively well-understood, the astrophysical production site of the $r$-process has been debated for 60 years \citep{Burbidge57,Cameron57}. Candidate sites must produce a strong neutron flux, so naturally, only the most violent events, such as supernova explosions or merging neutron stars can be considered. Moreoever, the low metallicity of $r$-process-enhanced stars requires a relatively fast enrichment, prior to their formation. This has been used \citep{sneden08,jacobson13} to argue for a  supernovae, as they provide prompt enrichment. Moreover, supernovae are of course also responsible for all the lighter fusion elements observed in metal-poor stars. But tension with theoretical models could not be resolved \citep{Arcones11,Wanajo13}. The recent discovery of the first $r$-process galaxy, Reticulum\,II, initiated a drastic re-interpretation of the astrophysics data and confirmed the importance of dwarf galaxy archaeology: The stars in this dwarf galaxy suggest that the $r$-process predominantly occurs in neutron star mergers. 

\subsection{The $r$-process galaxy Reticulum\,II}\label{ret}

\begin{figure}[!t]
\includegraphics[width=4.9in,clip=]{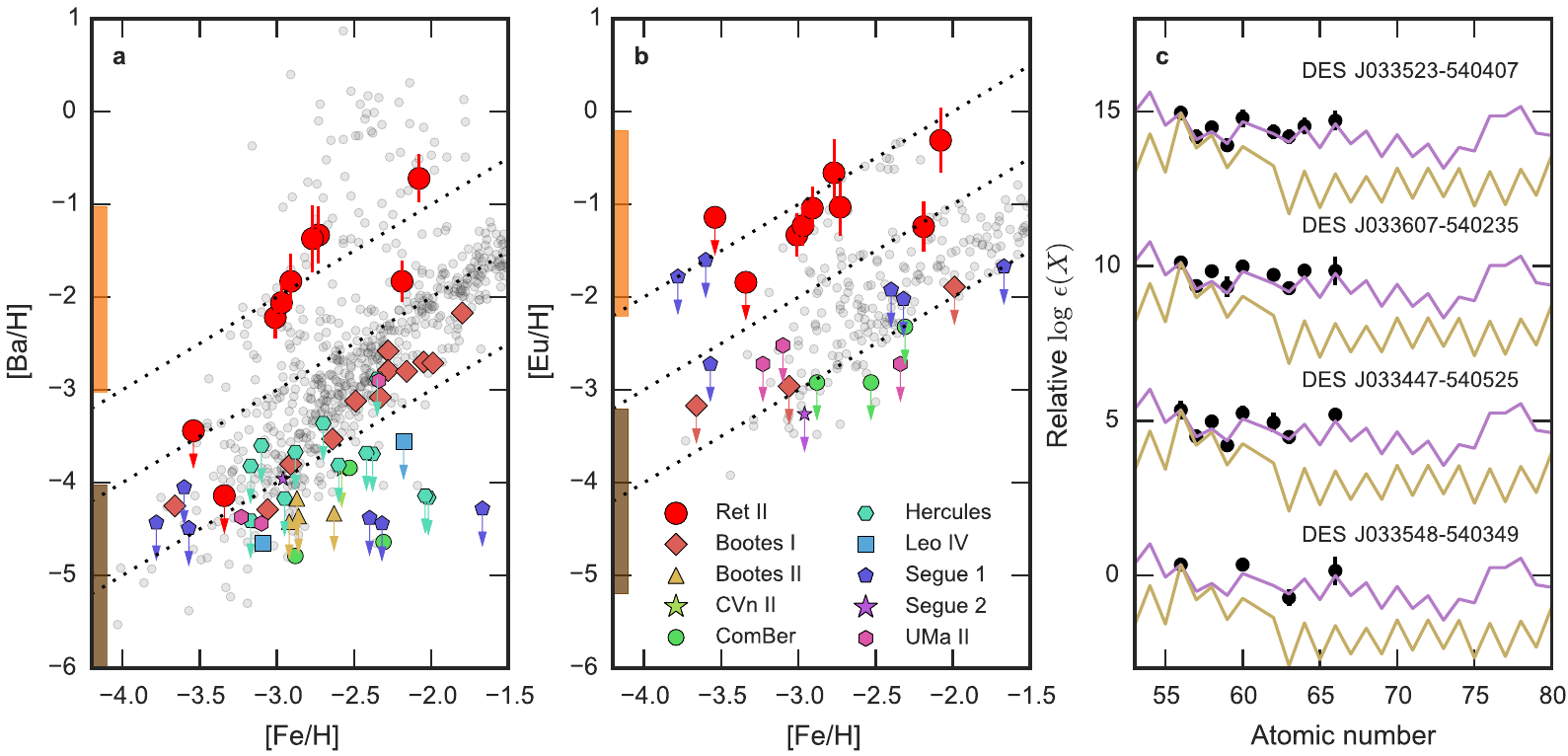}
  \caption{\label{ret_plot} Left: Barium and europium abundances of stars in Reticulum\,II (red points), in the halo (gray), and in ultra-faint dwarf galaxies (colored points) as functions of the iron abundance. Orange and brown vertical bars indicate yield ranges predicted by a neutron star merger and core-collapse supernovae, respectively. 
 Arrows denote upper limits. Right: Observed abundances overlaid with the scaled solar $r$- and $s$-process patterns in purple and gold, respectively. Figure from \citet{ji16b}.} 
\end{figure}

Reticulum\,II is a small, ultrafaint ($L \sim3,000$\,L$_\odot$)
dwarf galaxy located in the Galactic halo at only 30\,kpc distance from the Galactic center,
discovered in data from the Dark Energy Survey (DES) \citep{DrWag15b,Koposov15b}. Like all
other ultrafaint dwarfs, it contains only a few thousand stars,
but is highly dark matter dominated \citep{simon15}, remarkably
metal poor ($\mbox{[Fe/H]} = -2.65$), and extremely old (12 billion
years according to its color-magnitude diagram \citealt{Bechtol15}), and
devoid of gas. Chemical abundance studies based on high-resolution
spectroscopy have confirmed that stars in Reticulum\,II are indeed
chemically primitive \citep{ji16b, Roederer16b}, lending support to the idea that the ultrafaint dwarf galaxies have
preserved the signatures of the earliest metal-production events
in the Universe, and at least some of them may be first/early
galaxies that somehow survived to the present day
\citep{frebel12,frebel14,chiti18}. Furthermore, these
observations revealed that seven of the nine brightest
stars under investigation exhibit an unusual, extreme enhancement
in $r$-process elements (compared with most halo and all other dwarf
galaxy stars) and that their abundances match the scaled solar
$r$-process pattern above barium---they are all $r$-II stars (\textbf{Figure~\ref{ret_plot})}. Previously, this signature had only been observed in the
Milky Way $r$-process-enhanced halo stars, as described
above. Reticulum\,II is therefore the first known $r$-process
galaxy. It must have experienced a seemingly rare and prolific
$r$-process event, given that this signature was found only  in 1
of 10 ultrafaint dwarf galaxies that had at least one star
studied with high-resolution spectroscopy at that time. Note
that, since stars in the other nine ultrafaint dwarfs show
abundances of neutron-capture elements that are unusually
{low}, this finding was doubly surprising.

The enhanced abundances of elements associated with the main $r$-process clearly point to a neutron star merger as being responsible for the enrichment in Reticulum\,II \citep{ji16b,Ji16c, ji18}. Interestingly, prior to the discovery of Reticulum\,II, neutron star mergers were not believed to be suitable as main $r$-process-element producers, at least not in the early universe. At face value, the low metallicity of $r$-process-enhanced halo stars suggested a fast enrichment channel, more easily provided by supernovae. But this approach could not factor in any information on their birth environments, as the origins of halo stars are unknown. All stars in Reticulum\,II, however, come from the same dwarf galaxy. Consideration of environmental information, such as what might have occurred in a young Reticulum\,II, showed that about 100 million years are needed for the gas in such a small system to cool down sufficiently to form the next generation of stars \citep{BlandHaw15}, following a major energy injection by the explosion of the very first stars. Luckily, this time span appears long enough for a neutron star binary to in-spiral and merge \citep{Dominik12}. Adding up the yields of hundreds to thousands of supernovae, as an alternative, would instead cause disruption of the system. In addition, the gas mass of Reticulum\,II can be estimated, into which the yield of the $r$-process event was diluted into. A minimum gas mass of 10$^5$\,M$_\odot$ is given by the amount of swept up gas in a supernova explosion \citep{ji16b}, and also by the mass of typical star-forming clouds \citep{Ji15}. The total luminous (baryonic) mass of Reticulum\,II, 10$^7$\,M$_\odot$, provides an upper bound. From yield predictions of theoretical calculations of $r$-process nucleosynthesis occurring in neutron star mergers \citep{goriely11}, the expected abundance level for stars to form from the enriched gas was derived (assuming homogeneous and instantaneous metal mixing). The results agree well with the observed data for stars in Reticulum\,II. We note that the observational results would also agree with the yields from a magneto-rotationally driven jet supernovae, but recent results suggest these yields to be significantly lower than previously thought \citep{moesta17}. 

Considering Reticulum\,II, it should be added that of course some supernovae could have exploded, at least as part of the first generation of stars. A limited $r$-process likely took place, which provided small amounts of neutron-capture elements to the system (see Section~\ref{chemevol} for more details). In Reticulum\,II, such marginal enrichment could explain the nature of the two stars found that do not display $r$-process enhancement, but rather, extremely low neutron-capture element abundances. Sequential bursts of star formation might have produced the neutron-capture-poor stars first (they have slightly lower [Fe/H]) before the neutron star merger took place, after which the $r$-process-enhanced stars formed. Alternatively, inhomogeneous mixing might explain the abundance variations, or Reticulum\,II could have absorbed another small dwarf galaxy, which would imply that the neutron-capture-poor stars actually formed outside of Reticulum\,II. The latter scenario is appealing because the stars found in all other ultra-faint dwarf galaxies have equally low heavy-element abundances as these two stars.

Another aspect to consider is the light neutron-capture elements in Reticulum\,II stars. As can be seen at the bottom panel of \textbf{Figure~\ref{pattern_plot}}, these stars have some of the lowest and most discrepant first-peak abundances (Sr-Y-Zr) compared to the solar $r$-process pattern, when scaled to elements barium and higher. These deviations are further discussed for halo stars in Section~3.4.3, but it is interesting to specifically consider them in the context of Reticulum\,II. Recall that the main $r$-process leads to the production of second- and third-peak elements, but likely no first-peak elements. Since Reticulum\,II is the chemically cleanest environment yet found in which a neutron star merger occurred, the stable $r$-process pattern for elements barium and above can easily be explained this way (see also discussion above). What is observed in the first peak, however, must result from the supernovae that exploded in the system, with a limited $r$-process operating. The $r$-process-enhanced stars that exhibit large first-peak deviations, such as CS~22892-052 \citep{sneden2000}, might have originated from clean systems such as Reticulum\,II. Other $r$-process-enhanced stars with less-deviant abundances might correspondingly have formed from larger, more massive galaxies that underwent some level of chemical evolution, i.e., that experienced a greater number of supernovae. Overall, in the above scenario might explain the full range of variations among $r$-process-enhanced halo stars, including the solar pattern, given the variable number of supernovae that a host system experienced. While more theoretical studies are needed, these observations should thus principally be able to constrain the supernova population and the limited $r$-process. Interestingly, new observations \citep{ji18} have shown the brightest star in Reticulum\,II to be an actinide deficient star (see Section~\ref{rprocstars}), because it has a lower-than-expected thorium abundance. High-quality data could be obtained for this star, but the others are too faint. It will thus remain a question for some time whether the other $r$-process-enhanced stars in this galaxy also exhibit  actinide deficiency. This would provide an important constraint on $r$-process nucleosynthesis, including the question if another source or site of $r$-process elements is required to have occurred, or if the different components of a neutron star merger might ultimately explain this phenomenon.

Dwarf galaxy archaeology, i.e. observing stars in dwarf galaxies has one clear advantage. Given the known origin of these stars, the entire system can be theoretically modeled and thus better interpreted in terms of its chemical evolution and star-formation history. The clear downside is the huge amount of large-aperture telescope time required (typically one night per star) to observe these distant, and thus relatively faint, stars. Multi-fiber echelle spectrographs alleviate this issue somewhat by enabling observations of multiple (e.g. $\sim10$) stars at once but even then, only the very brightest star can be studied with current telescope and detector technology. Among $r$-process-enhanced halo stars, on the other hand, very bright metal-poor objects can be selected (itself a long-term effort), for which it is comparatively easier to obtain exquisite quality data, so that many elements, including uranium and lead, can be measured. But their disadvantage remains their unknown individual origins.

An overall comparison of $r$-process-enhanced halo stars with those found in Reticulum\,II shows that their absolute (that is, relative to H) enhancement levels match, suggesting a common origin in dwarf galaxies. Variations in the enhancement could simply be due to the amount gas present in the various dwarf galaxies into which $r$-process elements are diluted. Observational constraints from additional $r$-process dwarf galaxies as well as simulations of galaxy assembly should soon provide more information of how these two populations are linked. The discovery of Reticulum\,II underpins the importance of the astrophysics component, which is complementary to experimental nuclear physics efforts that obtain measurements, or very good predictions, of the fundamental properties (e.g., masses, nuclear interaction cross sections, and decay rates) of heavy neutron-rich nuclei. These efforts are a prime motivation for several international accelerator facilities, such as the Facility for Rare Isotope Beams (FRIB), now under construction at Michigan State University (expected completion in 2022), FAIR (Germany), RIKEN (Japan), and RAON (South Korea). Until many of the current nuclear physics uncertainties in $r$-process nucleosynthesis networks are resolved, interpretation of the abundance patterns of $r$-process-enhanced stars need to be carried out mindfully, so as not to ascribe any apparent discrepancies with models to properties of the astrophysical site.

\subsection{Other dwarf galaxies with $r$-process element signatures}\label{other_gals}

Soon after the discovery of Reticulum\,II, a second $r$-process galaxy was identified. The brightest star in the somewhat more-massive ultra-faint dwarf galaxy Tucana\,III was observed with high-resolution spectroscopy. It exhibits a moderate $r$-process enhancement, i.e., it is an $r$-I star \citep{Hansen17}. For elements above barium, the element pattern matches that of the scaled solar $r$-process, whereas first-peak elements deviate, as in the Reticulum\,II stars. Additional Tucana\,III stars also show similar enhancements (J. Marshall, personal comm.). Tucana\,III is believed to have been even more massive in the past, which might explain why its $r$-process level is lower than that found in Reticulum\,II, as any yield would likely have been more diluted.

A handful of $r$-II stars (classified based on the [Ba/Eu] ratio) were previously found in the 
more-massive classical dwarf galaxies Ursa Minor, Draco, and Fornax; see \citep{Hansen17} for more details. About 20 $r$-I stars also had been identified in Draco, Ursa Minor, Sculptor, Fornax and Carina \citep{shetrone03,cohen10,tsujimoto17}. While these stars do not all perfectly match the scaled solar $r$-process pattern (or simply do not have enough abundance measurements available to test for a match), witnessing signs of $r$-process nucleosynthesis implies that enrichments from neutron star mergers may be rare but not unique, and are not restricted to ultra-faint dwarf galaxies. The observed variations in absolute $r$-process enhancements in all of these systems might be due to different dilution masses, the timing of the neutron star merger, and subsequent chemical-evolution and star-formation history, and/or 
unique accretion histories. In this context, Draco is of particular interest. Its stars cluster around certain metallicities whereby each cluster shows distinctly different [Ba/Fe] ratios, possibly due to multiple events/processes that enriched Draco with time \citep{tsujimoto17}.

\subsection{The gravitational wave connection}

The LIGO and Virgo observatories recently detected, for the first time, the gravitational wave signal of a neutron star merger, GW170817 \citep{LIGOGW170817a,LIGOGW170817b}. The electromagnetic counterpart SSS17a was also observed \citep{Drout17,shappee17}, which is known as a ``kilonova'', and believed to be the radioactive decay afterglow following the production of unstable heavy neutron-rich isotopes during an $r$-process event. For stellar archaeology, the significance of this detection lies in the assumption that the $r$-process-enhanced stars in Reticulum\,II (and also other $r$-process-enhanced halo stars) are probing the ejecta of events like GW170817. The detailed abundances of $r$-process-enhanced stars provide important constraints on the nature of $r$-process event(s) that occurred in the early universe. This is complementary to gravitational wave science, which is now able to provide specific data on the astrophysical site of the $r$-process in the local universe, and to deep sea measurements of plutonium  deposited there by local $r$-process event(s) \citep{Wallner15}. 

Considering these two avenues, neutron star mergers thus appear to produce $r$-process elements throughout cosmic history. But several open questions remain to be addressed. The frequency of occurrence, both locally and in the early universe, is needed to incorporate neutron star mergers into chemical-evolution models, to broadly probe them being the dominant $r$-process-element producers. The currently inferred neutron star merger rate is $R_{\rm{NSM}}=1540^{+3200}_{-1220} $\,Gpc$^{-3}$ yr$^{-1}$ \citep{LIGOGW170817a}, for a core-collapse supernova rate of $R_{\rm{CCSN}} \approx 1.1 \pm 0.2 \times 10^{5}$\,Gpc$^{-3}$ yr$^{-1}$ \citep{Taylor14}. From Reticulum\,II, \citet{ji16b} estimated that one neutron star merger occurred for every ${\sim}2000$ core-collapse supernovae. Clearly, no exact agreement has yet been reached, as the GW17081-based values are a factor of $\sim10$ higher than what the Reticulum\,II case suggests. As more neutron star mergers and $r$-process dwarf galaxies are detected, these estimates will be improved.

The $r$-process yield of each event is also crucial to quantify. The total mass ejected by GW170817 is $\sim0.05$\,M$_\odot$. This may correspond to ${\sim}10^{-5}\,M_\odot$ of Eu per event \citep{Cote17}. This is not unlike what was found to match observations of the stars in Reticulum\,II, M$_{\rm Eu} \sim 10^{-4.5 \pm 1}$\,M$_\odot$ \citep{ji16b}. Nevertheless, yields need to be obtained for many neutron star mergers, to determine if there is a universal yield, or if individual properties of neutron star mergers and/or their binary and merger dynamics lead to yield variations. Modeling the $r$-process in neutron star mergers will then become much more detailed. However, theoretical yield calculations remain difficult, but could be dramatically aided by improved nuclear properties of the isotopes involved in $r$-process nucleosynthesis. New facilities such as FRIB will address these important shortcomings, as the results and implications are of interest to many subfields within astronomy and astrophysics. Along the way, model results will continue to be constrained with observations of $r$-process-enhanced halo stars and $r$-process galaxies to reach a more complete understanding of the origin of the heaviest elements.

\subsection{Chemical evolution of neutron-capture elements and the galaxy formation connection}\label{chemevol}
  
\begin{figure}[!t]
\includegraphics[width=4.5in]{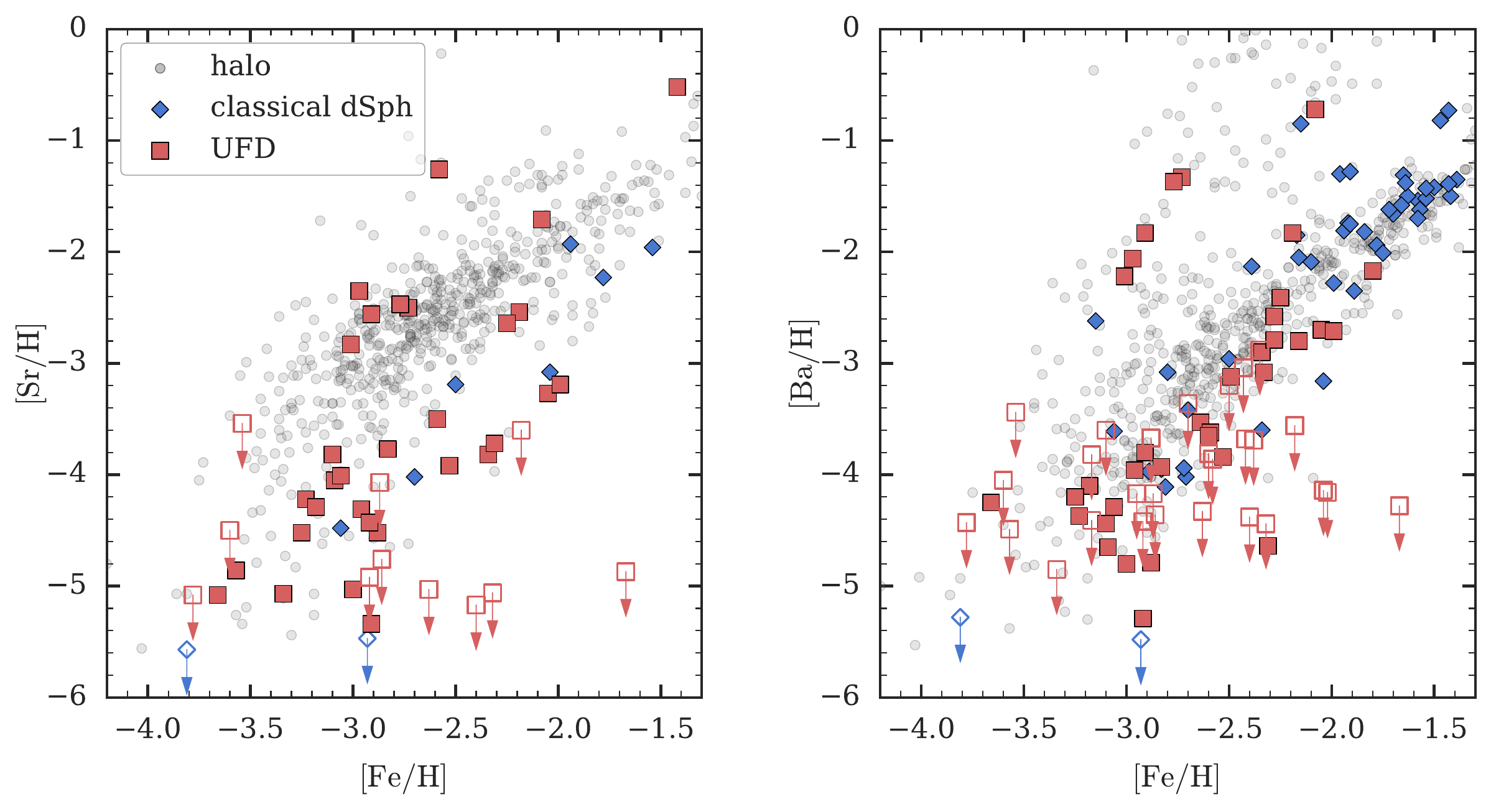}
\caption{\label{ncaps_plot} Evolution of the absolute abundances ([X/H]) of neutron-capture elements strontium and barium as a function of the iron abundance. Halo stars and stars from classical dwarf spheroidal and ultra-faint dwarf galaxies are shown. Upper limits are indicated with an arrow. Figure courtesy of A. Ji.}
\end{figure}

The $r$-process-enhanced stars only make up a small fraction of metal-poor stars \citep{heresII}. All other halo stars exhibit \citep{Roederer13} at least a low abundance of neutron-capture elements (typically measured for strontium and barium), although they do not show any characteristic abundance patterns that would point to a specific origin of these elements (see \citep{sneden08, jacobson13} for more details), as illustrated in \textbf{Figure~\ref{ncaps_plot}}. Generally, at the lowest metallicities, neutron-capture-element abundances are highly depleted. This indicates an early production of small amounts of these elements that were then contributed to the natal gas clouds of these stars. In particular, stars in the ultra-faint dwarf galaxies show extremely low abundances \citep{frebel14} (except those in $r$-process galaxies). This points to the limited $r$-process as a likely source, since it could operate either in any supernova and/or supernovae whose progenitors mass falls in a specific range \citep{lee_d_13}, at least at the earliest times. Rapidly rotating massive stars, however, might also produce these elements \citep{chiappini11}. The steep increase of, e.g., [Sr/Fe] at $\mbox{[Fe/H]}\sim-3$ points to an increase in neutron-capture elements independent of iron and other light elements. Then, from $\mbox{[Fe/H]}\sim-2.6$, the $s$-process contributes these elements as well, and the evolution of strontium and barium proceeds in lockstep with that of lighter elements. Theoretical modeling of the observed abundance trends of various neutron-capture elements has been challenging \citep{cescutti14}, but the addition of yields from neutron star mergers to Galactic chemical evolution models and hydrodynamical simulations of galaxy formation have proven very promising \citep{Tsujimoto14b, cescutti15, Cote17,naiman17}.

From a broader perspective, our galaxy formed from the accretion of smaller galaxies and building blocks, especially at the earliest times. The oldest stars found today in the Milky Way must date back to the earliest star-forming events, and thus likely originated in some of these small, early galaxies. These were later absorbed into the Milky Way, spilling all their stars into the outer region - the stellar halo - of our galaxy. As a consequence, any information on the original host galaxy, in which the stars formed, is lost. Applying this scenario to $r$-process-enhanced stars suggests that they may well have formed in now-accreted systems that were very similar to Reticulum\,II. After all, these halo stars are very ([Fe/H] $< -2$) and extremely ([Fe/H] $< -3$ metal-poor, and their $r$-process patterns and levels of enrichment strongly resemble those of the $r$-process-enhanced stars in Reticulum\,II. A common birth environment is thus likely. This implies that characteristic abundance patterns, such as that of the $r$-process, could be used to trace the hierarchical build-up of the Milky Way. More specifically, it might become possible to use the population of $r$-process-enhanced halo stars as tracers of the fraction of early accreted dwarf galaxies that built the halo of the Milky Way. This broadly follows the idea of chemical tagging, initially envisioned by \citep{blandhawthorn_freeman}, who suggested that the Galaxy's formation history is encoded in the chemical abundances of its stars.

To make this approach a reality, a census of all $r$-process-enhanced stars in both the Galaxy and dwarf galaxies needs to be undertaken. New observational efforts to identify additional $r$-process-enhanced halo stars are already underway \citep{Placco17,hansen18,sakari18,gull18}. Using a variety of telescopes in both hemispheres, over a dozen new $r$-II stars have been identified in the last 1-2 years alone; this effort is aimed at building up a sample of a total of $\sim100$ known $r$-II stars. Only in this way can statistically significant results be obtained that will provide a legacy data set for constraining theoretical and experimental $r$-process studies well into the era of experimental facilities such as FRIB. This follows initial efforts \citep{heresI} to increase the then-small number of known $r$-II stars to current numbers of $\sim30$ \citep{heresII, hayek09}, and to also deliver  several dozen $r$-I stars. Only a few $r$-II stars have been found from other efforts, e.g.,  \citep{he1523,lai2008,aoki10} 
Stars with $s$-process and $i$-process enhancement, or with a combined $r+s$ signature, have also been discovered by these efforts, although in less systematic ways. In addition, as new dwarf galaxies are being discovered, their brightest stars will be observed to check for $r$-process enrichment. Any new $r$-process galaxies are much needed to continue to establish the fraction of small systems that hosted a neutron star merger. This promising new avenue to study galaxy assembly with stellar chemical signatures is grounded in our current understanding of the nuclear physics of element production; only with concerted efforts on both the nuclear physics and astrophysics fronts can it reach its full potential.

\vspace{0.5cm}
The $r$-process-enhanced stars have previously been called the cosmic ``Rosetta Stone'' for deciphering the  origin of the $r$-process. This label should now be extended to the truly remarkable $r$-process galaxies, as they bring together seemingly different research areas from nuclear physics, to gravitational physics, to astrophysics in a most unique and fascinating way. This convergence, and the progress made by the gravitational wave detections, the new nuclear accelerator facilities (under construction), and the many theoretical and observational efforts, promise to finally resolve extremely challenging questions about the $r$-process. Increasingly strong constraints on the nature of the $r$-process and the identification of the astrophysical site of the $r$-process are imminent, and with it, a full understanding of the origin of the heaviest elements in the cosmos.



\begin{issues}[FUTURE ISSUES IN NUCLEAR ASTROPHYSICS]

Progress on the topics covered in this review can achieved if the nuclear physics and astrophysics communities jointly focus on their common goal: understanding the origin of the elements. 

\begin{enumerate}
\item Observational astrophysics needs to continue to deliver suitable metal-poor stars to study the various nucleosynthesis processes. Especially $r$-process-enhanced stars with detectable thorium \textit{and uranium} will provide the most stringent constraints on the $r$-process. Work on metal-poor stars in additional dwarf galaxies (e.g., those yet to be discovered with DES and LSST) will provide complementary information towards identifying the site(s) of the $r$-process.  Once the elemental yields are known, the gas dilution masses could be derived from stellar abundances. Such results would provide crucial constraints for cosmological simulations of early galaxy formation. Additional $s$-, $i$-, and $r+s$-process-enhanced stars will provide means to study the low to intermediate neutron-density regime, especially at low metallicity. This work needs to be combined with binary star mass-transfer modeling to correctly interpret the observed stellar abundances.

To ensure long term success, better optical and near-UV spectra of already known metal-poor stars are needed, along with improved atomic data, NLTE, and 3D stellar atmosphere modeling, as well as larger telescopes. Close collaboration between nuclear physics and stellar astronomy will be required to leverage the next-generation telescopes, such as the 25\,m Giant Magellan Telescope. When equipped with high-resolution spectrographs, they have the potential to provide currently missing data to move observational nuclear astrophysics into a new era of discovery.
 
\item Only nuclear physics theory can provide detailed yield predictions of the various nucleosynthesis processes. Much needed are low-metallicity $s$- and $i$-process yields, and $r$-process yields taking potential sites with different components into account (e.g., neutron star mergers). It also remains to be seen how much and which neutron-capture elements can be produced by ordinary core-collapse supernovae. These yields are crucial for interpreting the observational data. In parallel, if enough $r$-process-enhanced stars are found that can be assumed to be of old age, initial production ratios could be determined empirically, and compared with the theoretical predictions. 

\item Nuclear physics experiments, such as the forthcoming FRIB, need to continue exploring the properties of nuclei far from stability, to provide the fundamental physics inputs required for nucleosynthesis modeling. The success of nuclear astrophysics rests on our understanding of the basic properties of these nuclei.

\item Once all these achievements can be combined, it will be possible to finally understand where the $r$-process occurs in the universe, and why $r$-II stars seemingly all have the same [Fe/H] metallicity range, and no $r$-process-enhanced star has yet been found with $\mbox{[Fe/H]} <-3.5$. Comprehensively grasping the chemical evolution of $r$-process elements, and with the availability of low-metallicity $s$-process yields, the existence of CEMP-$r+s$ stars can be better  understand. Together with supernova yields for the limited $r$-process, this will provide constraints on the overall chemical evolution of neutron-capture elements. 

\end{enumerate}
\end{issues}

The authors are not aware of any affiliations, memberships, funding, or financial holdings that might be perceived as affecting the objectivity of this review. 

\section*{ACKNOWLEDGMENTS}

A.F. warmly thanks the JINA community for 15 years of fruitful discussions and enjoyable collaborations. She is indebted to Abdu Abohalima, Timothy Beers, Anirudh Chiti, Norbert Christlieb, Rana Ezzeddine, Alexander Ji, Amanda Karakas, Gail McLaughlin, John Norris, Vinicius Placco, Yong Qian, Ian Roederer, Hendrik Schatz, Joshua Simon, Rebecca Surman, and many others, for sharing the passion about the origins of the elements in cosmos. Kaley Brauer, Melanie Hampel, Alexander Ji, and Amanda Karakas kindly provided figures.
A.F. thanks the [Department of Energy's] Institute for Nuclear Theory at the University of Washington for hospitality during Program INT-17-2b and the Department of Energy for partial support during preparatory work of this review. A.F. thanks the participants of Program INT-17-2b for discussions that led to Table~2.
A.F. thanks the International Centre for Radio Astronomy Research (ICRAR) for hospitality during the final stages of the review. She acknowledges support from the ICRAR Visiting Fellowship For Senior Women In Astronomy, at University of Western Australia and Curtin University, Perth, Australia.
%



\bibliographystyle{Astronomy}
\bibliography{mybib,fn_araa,paper_sm0313,ref}

\begin{thebibliography}{}
\expandafter\ifx\csname natexlab\endcsname\relax\def\natexlab#1{#1}\fi

\bibitem[{{Abate} et~al.(2015){Abate}, {Pols}, {Izzard} \&
  {Karakas}}]{Abate2015-2}
{Abate} C, {Pols} OR, {Izzard} RG, {Karakas} AI. 2015.
\newblock \textit{\aap} 581:A22

\bibitem[{{Abbott} et~al.(2017{\natexlab{a}}){Abbott}, {Abbott}, {Abbott},
  {Acernese}, {Ackley} et~al.}]{LIGOGW170817a}
{Abbott} BP, {Abbott} R, {Abbott} TD, {Acernese} F, {Ackley} K, et~al.
  2017{\natexlab{a}}.
\newblock \textit{Physical Review Letters} 119:161101

\bibitem[{{Abbott} et~al.(2017{\natexlab{b}}){Abbott}, {Abbott}, {Abbott},
  {Acernese}, {Ackley} et~al.}]{LIGOGW170817b}
{Abbott} BP, {Abbott} R, {Abbott} TD, {Acernese} F, {Ackley} K, et~al.
  2017{\natexlab{b}}.
\newblock \textit{\apjl} 848:L12

\bibitem[{{Abia} et~al.(2002){Abia}, {Dom{\'{\i}}nguez}, {Gallino}, {Busso},
  {Masera} et~al.}]{abia02}
{Abia} C, {Dom{\'{\i}}nguez} I, {Gallino} R, {Busso} M, {Masera} S, et~al.
  2002.
\newblock \textit{\apj} 579:817

\bibitem[{{Abohalima} \& {Frebel}(2017)}]{abohalima17}
{Abohalima} A, {Frebel} A. 2017.
\newblock \textit{arXiv:1711.04410}

\bibitem[{{Aguado} et~al.(2018){Aguado}, {Gonz{\'a}lez Hern{\'a}ndez}, {Allende
  Prieto} \& {Rebolo}}]{aguado18}
{Aguado} DS, {Gonz{\'a}lez Hern{\'a}ndez} JI, {Allende Prieto} C, {Rebolo} R.
  2018.
\newblock \textit{\apjl} 852:L20

\bibitem[{{Amarsi} et~al.(2016){Amarsi}, {Lind}, {Asplund}, {Barklem} \&
  {Collet}}]{amarsi16}
{Amarsi} AM, {Lind} K, {Asplund} M, {Barklem} PS, {Collet} R. 2016.
\newblock \textit{\mnras} 463:1518

\bibitem[{{Aoki} et~al.(2007){Aoki}, {Beers}, {Christlieb}, {Norris}, {Ryan} \&
  {Tsangarides}}]{aoki07}
{Aoki} W, {Beers} TC, {Christlieb} N, {Norris} JE, {Ryan} SG, {Tsangarides} S.
  2007.
\newblock \textit{\apj} 655:492--521

\bibitem[{{Aoki} et~al.(2010){Aoki}, {Beers}, {Honda} \& {Carollo}}]{aoki10}
{Aoki} W, {Beers} TC, {Honda} S, {Carollo} D. 2010.
\newblock \textit{\apjl} 723:L201--L206

\bibitem[{{Aoki} et~al.(2001){Aoki}, {Ryan}, {Norris}, {Beers}, {Ando}
  et~al.}]{2001aokisprocess}
{Aoki} W, {Ryan} SG, {Norris} JE, {Beers} TC, {Ando} H, et~al. 2001.
\newblock \textit{ApJ} 561:346

\bibitem[{{Arcones} \& {Montes}(2011)}]{Arcones11}
{Arcones} A, {Montes} F. 2011.
\newblock \textit{\apj} 731:5

\bibitem[{{Arlandini} et~al.(1999){Arlandini}, {K{\"a}ppeler}, {Wisshak},
  {Gallino}, {Lugaro} et~al.}]{arlandini1999}
{Arlandini} C, {K{\"a}ppeler} F, {Wisshak} K, {Gallino} R, {Lugaro} M, et~al.
  1999.
\newblock \textit{ApJ} 525:886--900

\bibitem[{{Asplund}(2005)}]{asplund_araa}
{Asplund} M. 2005.
\newblock \textit{ARA\&A} 43:481

\bibitem[{{Barbuy} et~al.(2005){Barbuy}, {Spite}, {Spite}, {Hill}, {Cayrel}
  et~al.}]{barbuy2005}
{Barbuy} B, {Spite} M, {Spite} F, {Hill} V, {Cayrel} R, et~al. 2005.
\newblock \textit{A\&A} 429:1031

\bibitem[{{Barklem} et~al.(2005){Barklem}, {Christlieb}, {Beers}, {Hill},
  {Bessell} et~al.}]{heresII}
{Barklem} PS, {Christlieb} N, {Beers} TC, {Hill} V, {Bessell} MS, et~al. 2005.
\newblock \textit{A\&A} 439:129--151

\bibitem[{{Bechtol} et~al.(2015){Bechtol}, {Drlica-Wagner}, {Balbinot},
  {Pieres}, {Simon} et~al.}]{Bechtol15}
{Bechtol} K, {Drlica-Wagner} A, {Balbinot} E, {Pieres} A, {Simon} JD, et~al.
  2015.
\newblock \textit{\apj} 807:50

\bibitem[{{Beers} \& {Christlieb}(2005)}]{beers&christlieb05}
{Beers} TC, {Christlieb} N. 2005.
\newblock \textit{\araa} 43:531--580

\bibitem[{{Bergemann} et~al.(2012){Bergemann}, {Lind}, {Collet}, {Magic} \&
  {Asplund}}]{bergemann12}
{Bergemann} M, {Lind} K, {Collet} R, {Magic} Z, {Asplund} M. 2012.
\newblock \textit{MNRAS} 427:27

\bibitem[{{Bisterzo} et~al.(2010){Bisterzo}, {Gallino}, {Straniero},
  {Cristallo} \& {K{\"a}ppeler}}]{bisterzo10}
{Bisterzo} S, {Gallino} R, {Straniero} O, {Cristallo} S, {K{\"a}ppeler} F.
  2010.
\newblock \textit{\mnras} 404:1529--1544

\bibitem[{{Bisterzo} et~al.(2012){Bisterzo}, {Gallino}, {Straniero},
  {Cristallo} \& {K{\"a}ppeler}}]{Bisterzo12}
{Bisterzo} S, {Gallino} R, {Straniero} O, {Cristallo} S, {K{\"a}ppeler} F.
  2012.
\newblock \textit{\mnras} 422:849

\bibitem[{{Bisterzo} et~al.(2017){Bisterzo}, {Travaglio}, {Wiescher},
  {K{\"a}ppeler} \& {Gallino}}]{bisterzo17}
{Bisterzo} S, {Travaglio} C, {Wiescher} M, {K{\"a}ppeler} F, {Gallino} R. 2017.
\newblock \textit{\apj} 835:97

\bibitem[{{Bland-Hawthorn} \& {Gerhard}(2016)}]{bland_haw16}
{Bland-Hawthorn} J, {Gerhard} O. 2016.
\newblock \textit{\araa} 54:529

\bibitem[{{Bland-Hawthorn}, {Sutherland} \& {Webster}(2015)}]{BlandHaw15}
{Bland-Hawthorn} J, {Sutherland} R, {Webster} D. 2015.
\newblock \textit{\apj} 807:154

\bibitem[{{Bromm}, {Coppi} \& {Larson}(2002)}]{bromm02}
{Bromm} V, {Coppi} PS, {Larson} RB. 2002.
\newblock \textit{ApJ} 564:23--51

\bibitem[{{Burbidge} et~al.(1957){Burbidge}, {Burbidge}, {Fowler} \&
  {Hoyle}}]{Burbidge57}
{Burbidge} EM, {Burbidge} GR, {Fowler} WA, {Hoyle} F. 1957.
\newblock \textit{Reviews of Modern Physics} 29:547--650

\bibitem[{{Burris} et~al.(2000){Burris}, {Pilachowski}, {Armandroff}, {Sneden},
  {Cowan} \& {Roe}}]{Burris00}
{Burris} DL, {Pilachowski} CA, {Armandroff} TE, {Sneden} C, {Cowan} JJ, {Roe}
  H. 2000.
\newblock \textit{\apj} 544:302

\bibitem[{{Busso}, {Gallino} \& {Wasserburg}(1999)}]{busso_gallino_AGB1999}
{Busso} M, {Gallino} R, {Wasserburg} GJ. 1999.
\newblock \textit{AR\&A} 37:239

\bibitem[{{Caffau} et~al.(2011){Caffau}, {Bonifacio}, {Fran{\c c}ois},
  {Sbordone}, {Monaco} et~al.}]{caffau11}
{Caffau} E, {Bonifacio} P, {Fran{\c c}ois} P, {Sbordone} L, {Monaco} L, et~al.
  2011.
\newblock \textit{Nature} 477:67--69

\bibitem[{{Cameron}(1957)}]{Cameron57}
{Cameron} AGW. 1957.
\newblock \textit{\pasp} 69:201

\bibitem[{{Cameron}(1958)}]{cameron58}
{Cameron} AGW. 1958.
\newblock \textit{Annual Review of Nuclear and Particle Science} 8:299--326

\bibitem[{{Campbell} \& {Lattanzio}(2008)}]{campbell08}
{Campbell} SW, {Lattanzio} JC. 2008.
\newblock \textit{\aap} 490:769

\bibitem[{{Campbell}, {Lugaro} \& {Karakas}(2010)}]{campbell10}
{Campbell} SW, {Lugaro} M, {Karakas} AI. 2010.
\newblock \textit{\aap} 522:L6

\bibitem[{{Carney} et~al.(2003){Carney}, {Latham}, {Stefanik}, {Laird} \&
  {Morse}}]{carney2003}
{Carney} BW, {Latham} DW, {Stefanik} RP, {Laird} JB, {Morse} JA. 2003.
\newblock \textit{AJ} 125:293

\bibitem[{Cayrel et~al.(2001)Cayrel, Hill, Beers, Barbuy, Spite
  et~al.}]{Cayreletal:2001}
Cayrel R, Hill V, Beers T, Barbuy B, Spite M, et~al. 2001.
\newblock \textit{Nature} 409:691--692

\bibitem[{{Cescutti} \& {Chiappini}(2014)}]{cescutti14}
{Cescutti} G, {Chiappini} C. 2014.
\newblock \textit{\aap} 565:A51

\bibitem[{{Cescutti} et~al.(2015){Cescutti}, {Romano}, {Matteucci}, {Chiappini}
  \& {Hirschi}}]{cescutti15}
{Cescutti} G, {Romano} D, {Matteucci} F, {Chiappini} C, {Hirschi} R. 2015.
\newblock \textit{\aap} 577:A139

\bibitem[{{Chiappini} et~al.(2011){Chiappini}, {Frischknecht}, {Meynet},
  {Hirschi}, {Barbuy} et~al.}]{chiappini11}
{Chiappini} C, {Frischknecht} U, {Meynet} G, {Hirschi} R, {Barbuy} B, et~al.
  2011.
\newblock \textit{Nature} 472:454

\bibitem[{{Chiti} et~al.(2018){Chiti}, {Frebel}, {Ji}, {Jerjen}, {Kim} \&
  {Norris}}]{chiti18}
{Chiti} A, {Frebel} A, {Ji} AP, {Jerjen} H, {Kim} D, {Norris} JE. 2018.
\newblock \textit{\apj} 857:74

\bibitem[{{Christlieb} et~al.(2004{\natexlab{a}}){Christlieb}, {Beers},
  {Barklem}, {Bessell}, {Hill} et~al.}]{heresI}
{Christlieb} N, {Beers} TC, {Barklem} PS, {Bessell} M, {Hill} V, et~al.
  2004{\natexlab{a}}.
\newblock \textit{A\&A} 428:1027

\bibitem[{{Christlieb} et~al.(2004{\natexlab{b}}){Christlieb}, {Gustafsson},
  {Korn}, {Barklem}, {Beers} et~al.}]{christliebetal04}
{Christlieb} N, {Gustafsson} B, {Korn} AJ, {Barklem} PS, {Beers} TC, et~al.
  2004{\natexlab{b}}.
\newblock \textit{\apj} 603:708--728

\bibitem[{{Cohen} et~al.(2003){Cohen}, {Christlieb}, {Qian} \&
  {Wasserburg}}]{cohen2003}
{Cohen} JG, {Christlieb} N, {Qian} YZ, {Wasserburg} GJ. 2003.
\newblock \textit{ApJ} 588:1082--1098

\bibitem[{{Cohen} \& {Huang}(2010)}]{cohen10}
{Cohen} JG, {Huang} W. 2010.
\newblock \textit{\apj} 719:931--949

\bibitem[{{C{\^o}t{\'e}} et~al.(2017){C{\^o}t{\'e}}, {Fryer}, {Belczynski},
  {Korobkin}, {Chru{\'s}li{\'n}ska} et~al.}]{Cote17}
{C{\^o}t{\'e}} B, {Fryer} CL, {Belczynski} K, {Korobkin} O,
  {Chru{\'s}li{\'n}ska} M, et~al. 2017.
\newblock \textit{arXiv:1710.05875}

\bibitem[{{Cowan} \& {Rose}(1977)}]{cowan_rose77}
{Cowan} JJ, {Rose} WK. 1977.
\newblock \textit{\apj} 212:149

\bibitem[{{Cowan} et~al.(2002){Cowan}, {Sneden}, {Burles}, {Ivans}, {Beers}
  et~al.}]{cowan_U_02}
{Cowan} JJ, {Sneden} C, {Burles} S, {Ivans} II, {Beers} TC, et~al. 2002.
\newblock \textit{ApJ} 572:861

\bibitem[{{Cristallo} et~al.(2015){Cristallo}, {Straniero}, {Piersanti} \&
  {Gobrecht}}]{cristallo15}
{Cristallo} S, {Straniero} O, {Piersanti} L, {Gobrecht} D. 2015.
\newblock \textit{\apjs} 219:40

\bibitem[{{Dardelet} et~al.(2015){Dardelet}, {Ritter}, {Prado}, {Heringer},
  {Higgs} et~al.}]{dardelet15}
{Dardelet} L, {Ritter} C, {Prado} P, {Heringer} E, {Higgs} C, et~al. 2015.
\newblock \textit{arXiv:1505.05500}

\bibitem[{{de Mink} \& {Belczynski}(2015)}]{demink15}
{de Mink} SE, {Belczynski} K. 2015.
\newblock \textit{\apj} 814:58

\bibitem[{{Denissenkov} et~al.(2017){Denissenkov}, {Herwig}, {Battino},
  {Ritter}, {Pignatari} et~al.}]{denissenkov17}
{Denissenkov} PA, {Herwig} F, {Battino} U, {Ritter} C, {Pignatari} M, et~al.
  2017.
\newblock \textit{\apjl} 834:L10

\bibitem[{{Doherty} et~al.(2015){Doherty}, {Gil-Pons}, {Siess}, {Lattanzio} \&
  {Lau}}]{doherty15}
{Doherty} CL, {Gil-Pons} P, {Siess} L, {Lattanzio} JC, {Lau} HHB. 2015.
\newblock \textit{\mnras} 446:2599--2612

\bibitem[{{Dominik} et~al.(2012){Dominik}, {Belczynski}, {Fryer}, {Holz},
  {Berti} et~al.}]{Dominik12}
{Dominik} M, {Belczynski} K, {Fryer} C, {Holz} DE, {Berti} E, et~al. 2012.
\newblock \textit{\apj} 759:52

\bibitem[{{Drlica-Wagner} et~al.(2015){Drlica-Wagner}, {Bechtol}, {Rykoff},
  {Luque}, {Queiroz} et~al.}]{DrWag15b}
{Drlica-Wagner} A, {Bechtol} K, {Rykoff} ES, {Luque} E, {Queiroz} A, et~al.
  2015.
\newblock \textit{\apj} 813:109

\bibitem[{{Drout} et~al.(2017){Drout}, {Piro}, {Shappee}, {Kilpatrick}, {Simon}
  et~al.}]{Drout17}
{Drout} MR, {Piro} AL, {Shappee} BJ, {Kilpatrick} CD, {Simon} JD, et~al. 2017.
\newblock \textit{arXiv:1710.05443}

\bibitem[{{Eichler} et~al.(2015){Eichler}, {Arcones}, {Kelic}, {Korobkin},
  {Langanke} et~al.}]{Eichler15}
{Eichler} M, {Arcones} A, {Kelic} A, {Korobkin} O, {Langanke} K, et~al. 2015.
\newblock \textit{\apj} 808:30

\bibitem[{{Ezzeddine} et~al.(2016){Ezzeddine}, {Plez}, {Merle}, {Gebran} \&
  {Th{\'e}venin}}]{ezzeddine16}
{Ezzeddine} R, {Plez} B, {Merle} T, {Gebran} M, {Th{\'e}venin} F. 2016.
\newblock \textit{arXiv:1612.09302}

\bibitem[{{Feltzing} \& {Gonzalez}(2001)}]{feltzing01}
{Feltzing} S, {Gonzalez} G. 2001.
\newblock \textit{\aap} 367:253--265

\bibitem[{{Fern{\'a}ndez} \& {Metzger}(2016)}]{fernandez16}
{Fern{\'a}ndez} R, {Metzger} BD. 2016.
\newblock \textit{Annual Review of Nuclear and Particle Science} 66:23

\bibitem[{{Frebel}(2010)}]{frebel10_AN}
{Frebel} A. 2010.
\newblock \textit{Astronomische Nachrichten} 331:474

\bibitem[{{Frebel} et~al.(2005){Frebel}, {Aoki}, {Christlieb}, {Ando},
  {Asplund} et~al.}]{HE1327_Nature}
{Frebel} A, {Aoki} W, {Christlieb} N, {Ando} H, {Asplund} M, et~al. 2005.
\newblock \textit{Nature} 434:871--873

\bibitem[{{Frebel} \& {Bromm}(2012)}]{frebel12}
{Frebel} A, {Bromm} V. 2012.
\newblock \textit{\apj} 759:115

\bibitem[{{Frebel} et~al.(2007){Frebel}, {Christlieb}, {Norris}, {Thom},
  {Beers} \& {Rhee}}]{he1523}
{Frebel} A, {Christlieb} N, {Norris} JE, {Thom} C, {Beers} TC, {Rhee} J. 2007.
\newblock \textit{ApJL} 660:L117

\bibitem[{{Frebel} et~al.(2008){Frebel}, {Collet}, {Eriksson}, {Christlieb} \&
  {Aoki}}]{frebel08}
{Frebel} A, {Collet} R, {Eriksson} K, {Christlieb} N, {Aoki} W. 2008.
\newblock \textit{\apj} 684:588--602

\bibitem[{{Frebel}, {Johnson} \& {Bromm}(2007)}]{frebel07}
{Frebel} A, {Johnson} JL, {Bromm} V. 2007.
\newblock \textit{\mnras} 380:L40--L44

\bibitem[{{Frebel} \& {Norris}(2013)}]{psss}
{Frebel} A, {Norris} JE. 2013.
\newblock \textit{{Metal-Poor Stars and the Chemical Enrichment of the
  Universe}}.
\newblock ~55

\bibitem[{{Frebel} \& {Norris}(2015)}]{fn15}
{Frebel} A, {Norris} JE. 2015.
\newblock \textit{\araa} 53:631--688

\bibitem[{{Frebel}, {Simon} \& {Kirby}(2014)}]{frebel14}
{Frebel} A, {Simon} JD, {Kirby} EN. 2014.
\newblock \textit{\apj} 786:74

\bibitem[{{Freeman} \& {Bland-Hawthorn}(2002)}]{blandhawthorn_freeman}
{Freeman} K, {Bland-Hawthorn} J. 2002.
\newblock \textit{ARA\&A} 40:487

\bibitem[{{Frischknecht}, {Hirschi} \& {Thielemann}(2012)}]{frischknecht12}
{Frischknecht} U, {Hirschi} R, {Thielemann} FK. 2012.
\newblock \textit{\aap} 538:L2

\bibitem[{{Fujiya} et~al.(2013){Fujiya}, {Hoppe}, {Zinner}, {Pignatari} \&
  {Herwig}}]{fujiya13}
{Fujiya} W, {Hoppe} P, {Zinner} E, {Pignatari} M, {Herwig} F. 2013.
\newblock \textit{\apjl} 776:L29

\bibitem[{{Gallagher} et~al.(2016){Gallagher}, {Caffau}, {Bonifacio}, {Ludwig},
  {Steffen} \& {Spite}}]{gallagher16}
{Gallagher} AJ, {Caffau} E, {Bonifacio} P, {Ludwig} HG, {Steffen} M, {Spite} M.
  2016.
\newblock \textit{\aap} 593:A48

\bibitem[{{Gallagher} et~al.(2012){Gallagher}, {Ryan}, {Hosford},
  {Garc{\'{\i}}a P{\'e}rez}, {Aoki} \& {Honda}}]{gallagher12}
{Gallagher} AJ, {Ryan} SG, {Hosford} A, {Garc{\'{\i}}a P{\'e}rez} AE, {Aoki} W,
  {Honda} S. 2012.
\newblock \textit{\aap} 538:A118

\bibitem[{{Gallino} et~al.(1998){Gallino}, {Arlandini}, {Busso}, {Lugaro},
  {Travaglio} et~al.}]{gallino1998}
{Gallino} R, {Arlandini} C, {Busso} M, {Lugaro} M, {Travaglio} C, et~al. 1998.
\newblock \textit{ApJ} 497:388

\bibitem[{{Garc{\'{\i}}a P{\'e}rez} et~al.(2016){Garc{\'{\i}}a P{\'e}rez},
  {Allende Prieto}, {Holtzman}, {Shetrone}, {M{\'e}sz{\'a}ros}
  et~al.}]{garciaperez16}
{Garc{\'{\i}}a P{\'e}rez} AE, {Allende Prieto} C, {Holtzman} JA, {Shetrone} M,
  {M{\'e}sz{\'a}ros} S, et~al. 2016.
\newblock \textit{\aj} 151:144

\bibitem[{{Goriely}, {Bauswein} \& {Janka}(2011)}]{goriely11}
{Goriely} S, {Bauswein} A, {Janka} HT. 2011.
\newblock \textit{\apjl} 738:L32

\bibitem[{{Goriely} \& {Mowlavi}(2000)}]{goriely00}
{Goriely} S, {Mowlavi} N. 2000.
\newblock \textit{\aap} 362:599

\bibitem[{{Gull} et~al.(2018){Gull}, {Frebel}, {Cain}, {Placco}, {Ji}
  et~al.}]{gull18}
{Gull} M, {Frebel} A, {Cain} MG, {Placco} VM, {Ji} AP, et~al. 2018.
\newblock \textit{arXiv:1806.00645}

\bibitem[{{Hampel} et~al.(2016){Hampel}, {Stancliffe}, {Lugaro} \&
  {Meyer}}]{hampel16}
{Hampel} M, {Stancliffe} RJ, {Lugaro} M, {Meyer} BS. 2016.
\newblock \textit{\apj} 831:171

\bibitem[{{Hansen} et~al.(2016){Hansen}, {Andersen}, {Nordstr{\"o}m}, {Beers},
  {Placco} et~al.}]{hansen16_sproc_bin}
{Hansen} TT, {Andersen} J, {Nordstr{\"o}m} B, {Beers} TC, {Placco} VM, et~al.
  2016.
\newblock \textit{\aap} 588:A3

\bibitem[{{Hansen} et~al.(2015){Hansen}, {Andersen}, {Nordstr{\"o}m}, {Beers},
  {Yoon} \& {Buchhave}}]{hansen15_rproc_bin}
{Hansen} TT, {Andersen} J, {Nordstr{\"o}m} B, {Beers} TC, {Yoon} J, {Buchhave}
  LA. 2015.
\newblock \textit{\aap} 583:A49

\bibitem[{{Hansen} et~al.(2018){Hansen}, {Holmbeck}, {Beers}, {Placco},
  {Roederer} et~al.}]{hansen18}
{Hansen} TT, {Holmbeck} EM, {Beers} TC, {Placco} VM, {Roederer} IU, et~al.
  2018.
\newblock \textit{arXiv:1804.03114}

\bibitem[{{Hansen} et~al.(2017){Hansen}, {Simon}, {Marshall}, {Li}, {Carollo}
  et~al.}]{Hansen17}
{Hansen} TT, {Simon} JD, {Marshall} JL, {Li} TS, {Carollo} D, et~al. 2017.
\newblock \textit{\apj} 838:44

\bibitem[{{Hayek} et~al.(2009){Hayek}, {Wiesendahl}, {Christlieb}, {Eriksson},
  {Korn} et~al.}]{hayek09}
{Hayek} W, {Wiesendahl} U, {Christlieb} N, {Eriksson} K, {Korn} AJ, et~al.
  2009.
\newblock \textit{\aap} 504:511--524

\bibitem[{{Herwig}(2005)}]{herwig05}
{Herwig} F. 2005.
\newblock \textit{\araa} 43:435

\bibitem[{{Herwig} et~al.(2011){Herwig}, {Pignatari}, {Woodward}, {Porter},
  {Rockefeller} et~al.}]{herwig11}
{Herwig} F, {Pignatari} M, {Woodward} PR, {Porter} DH, {Rockefeller} G, et~al.
  2011.
\newblock \textit{\apj} 727:89

\bibitem[{{Hill} et~al.(2017){Hill}, {Christlieb}, {Beers}, {Barklem}, {Kratz}
  et~al.}]{Hill17}
{Hill} V, {Christlieb} N, {Beers} TC, {Barklem} PS, {Kratz} KL, et~al. 2017.
\newblock \textit{\aap} 607:A91

\bibitem[{{Hill} et~al.(2002){Hill}, {Plez}, {Cayrel}, {Beers}, {Nordstr{\"o}m}
  et~al.}]{Hill02}
{Hill} V, {Plez} B, {Cayrel} R, {Beers} TC, {Nordstr{\"o}m} B, et~al. 2002.
\newblock \textit{\aap} 387:560--579

\bibitem[{{Honda} et~al.(2006){Honda}, {Aoki}, {Ishimaru}, {Wanajo} \&
  {Ryan}}]{honda06}
{Honda} S, {Aoki} W, {Ishimaru} Y, {Wanajo} S, {Ryan} SG. 2006.
\newblock \textit{\apj} 643:1180--1189

\bibitem[{{Ivans} et~al.(2005){Ivans}, {Sneden}, {Gallino}, {Cowan} \&
  {Preston}}]{ivans05}
{Ivans} II, {Sneden} C, {Gallino} R, {Cowan} JJ, {Preston} GW. 2005.
\newblock \textit{\apjl} 627:L145--L148

\bibitem[{{Jacobson} \& {Frebel}(2014)}]{jacobson13}
{Jacobson} HR, {Frebel} A. 2014.
\newblock \textit{Journal of Physics G Nuclear Physics} 41:044001

\bibitem[{{Ji} \& {Frebel}(2018)}]{ji18}
{Ji} AP, {Frebel} A. 2018.
\newblock \textit{\apj} 856:138

\bibitem[{{Ji}, {Frebel} \& {Bromm}(2015)}]{Ji15}
{Ji} AP, {Frebel} A, {Bromm} V. 2015.
\newblock \textit{\mnras} 454:659

\bibitem[{{Ji} et~al.(2016{\natexlab{a}}){Ji}, {Frebel}, {Chiti} \&
  {Simon}}]{ji16b}
{Ji} AP, {Frebel} A, {Chiti} A, {Simon} JD. 2016{\natexlab{a}}.
\newblock \textit{\nat} 531:610

\bibitem[{{Ji} et~al.(2016{\natexlab{b}}){Ji}, {Frebel}, {Simon} \&
  {Chiti}}]{Ji16c}
{Ji} AP, {Frebel} A, {Simon} JD, {Chiti} A. 2016{\natexlab{b}}.
\newblock \textit{\apj} 830:93

\bibitem[{{Johnson} \& {Bolte}(2001)}]{johnson_bolte}
{Johnson} JA, {Bolte} M. 2001.
\newblock \textit{ApJ} 554:888

\bibitem[{{Jones} et~al.(2016){Jones}, {Ritter}, {Herwig}, {Fryer}, {Pignatari}
  et~al.}]{jones16}
{Jones} S, {Ritter} C, {Herwig} F, {Fryer} C, {Pignatari} M, et~al. 2016.
\newblock \textit{\mnras} 455:3848

\bibitem[{{Jonsell} et~al.(2006){Jonsell}, {Barklem}, {Gustafsson},
  {Christlieb}, {Hill} et~al.}]{jonsell06}
{Jonsell} K, {Barklem} PS, {Gustafsson} B, {Christlieb} N, {Hill} V, et~al.
  2006.
\newblock \textit{A\&A} 451:651

\bibitem[{{Karakas}(2010)}]{karakas11_models}
{Karakas} AI. 2010.
\newblock \textit{VizieR Online Data Catalog} 740:31413

\bibitem[{{Karakas}, {Garc{\'{\i}}a-Hern{\'a}ndez} \&
  {Lugaro}(2012)}]{karakas12}
{Karakas} AI, {Garc{\'{\i}}a-Hern{\'a}ndez} DA, {Lugaro} M. 2012.
\newblock \textit{\apj} 751:8

\bibitem[{{Karakas} \& {Lattanzio}(2014)}]{karakas14}
{Karakas} AI, {Lattanzio} JC. 2014.
\newblock \textit{PASA} 31:e030

\bibitem[{{Keller} et~al.(2014){Keller}, {Bessell}, {Frebel}, {Casey},
  {Asplund} et~al.}]{keller14}
{Keller} SC, {Bessell} MS, {Frebel} A, {Casey} AR, {Asplund} M, et~al. 2014.
\newblock \textit{Nature} 506:463

\bibitem[{{Koposov} et~al.(2015){Koposov}, {Casey}, {Belokurov}, {Lewis},
  {Gilmore} et~al.}]{Koposov15b}
{Koposov} SE, {Casey} AR, {Belokurov} V, {Lewis} JR, {Gilmore} G, et~al. 2015.
\newblock \textit{\apj} 811:62

\bibitem[{{Korobkin} et~al.(2012){Korobkin}, {Rosswog}, {Arcones} \&
  {Winteler}}]{Korobkin12}
{Korobkin} O, {Rosswog} S, {Arcones} A, {Winteler} C. 2012.
\newblock \textit{\mnras} 426:1940--1949

\bibitem[{{Kratz} et~al.(2007){Kratz}, {Farouqi}, {Pfeiffer}, {Truran},
  {Sneden} \& {Cowan}}]{Kratz07}
{Kratz} KL, {Farouqi} K, {Pfeiffer} B, {Truran} JW, {Sneden} C, {Cowan} JJ.
  2007.
\newblock \textit{\apj} 662:39--52

\bibitem[{{Kratz} et~al.(2004){Kratz}, {Pfeiffer}, {Cowan} \&
  {Sneden}}]{kratz2004}
{Kratz} KL, {Pfeiffer} B, {Cowan} JJ, {Sneden} C. 2004.
\newblock \textit{New Astronomy Review} 48:105

\bibitem[{{Lai} et~al.(2008){Lai}, {Bolte}, {Johnson}, {Lucatello}, {Heger} \&
  {Woosley}}]{lai2008}
{Lai} DK, {Bolte} M, {Johnson} JA, {Lucatello} S, {Heger} A, {Woosley} SE.
  2008.
\newblock \textit{ApJ} 681:1524

\bibitem[{{Lee} et~al.(2013){Lee}, {Johnston}, {Tumlinson}, {Sen} \&
  {Simon}}]{lee_d_13}
{Lee} DM, {Johnston} KV, {Tumlinson} J, {Sen} B, {Simon} JD. 2013.
\newblock \textit{Ap.J.} 774:103

\bibitem[{{Lind}, {Bergemann} \& {Asplund}(2012)}]{lind12}
{Lind} K, {Bergemann} M, {Asplund} M. 2012.
\newblock \textit{MNRAS} 427:50

\bibitem[{{Lippuner} \& {Roberts}(2015)}]{Lippuner15}
{Lippuner} J, {Roberts} LF. 2015.
\newblock \textit{\apj} 815:82

\bibitem[{{Lugaro}, {Campbell} \& {de Mink}(2009)}]{lugaro09}
{Lugaro} M, {Campbell} SW, {de Mink} SE. 2009.
\newblock \textit{PASA} 26:322

\bibitem[{{Lugaro} et~al.(2012){Lugaro}, {Karakas}, {Stancliffe} \&
  {Rijs}}]{lugaro12}
{Lugaro} M, {Karakas} AI, {Stancliffe} RJ, {Rijs} C. 2012.
\newblock \textit{\apj} 747:2

\bibitem[{{Mashonkina} et~al.(2010){Mashonkina}, {Christlieb}, {Barklem},
  {Hill}, {Beers} \& {Velichko}}]{heres5}
{Mashonkina} L, {Christlieb} N, {Barklem} PS, {Hill} V, {Beers} TC, {Velichko}
  A. 2010.
\newblock \textit{A\&A} 516:46

\bibitem[{{Mashonkina} et~al.(2011){Mashonkina}, {Gehren}, {Shi}, {Korn} \&
  {Grupp}}]{mashonkina11}
{Mashonkina} L, {Gehren} T, {Shi} JR, {Korn} AJ, {Grupp} F. 2011.
\newblock \textit{\aap} 528:A87

\bibitem[{{Metzger} et~al.(2010){Metzger}, {Mart{\'{\i}}nez-Pinedo}, {Darbha},
  {Quataert}, {Arcones} et~al.}]{Metzger10}
{Metzger} BD, {Mart{\'{\i}}nez-Pinedo} G, {Darbha} S, {Quataert} E, {Arcones}
  A, et~al. 2010.
\newblock \textit{\mnras} 406:2650--2662

\bibitem[{{Meyer}, {Krishnan} \& {Clayton}(1998)}]{meyer98}
{Meyer} BS, {Krishnan} TD, {Clayton} DD. 1998.
\newblock \textit{\apj} 498:808

\bibitem[{{M{\"o}sta} et~al.(2017){M{\"o}sta}, {Roberts}, {Halevi}, {Ott},
  {Lippuner} et~al.}]{moesta17}
{M{\"o}sta} P, {Roberts} LF, {Halevi} G, {Ott} CD, {Lippuner} J, et~al. 2017.
\newblock \textit{arXiv:1712.09370}

\bibitem[{{Naiman} et~al.(2017){Naiman}, {Pillepich}, {Springel},
  {Ramirez-Ruiz}, {Torrey} et~al.}]{naiman17}
{Naiman} JP, {Pillepich} A, {Springel} V, {Ramirez-Ruiz} E, {Torrey} P, et~al.
  2017.
\newblock \textit{arXiv:1707.03401}

\bibitem[{{Ness} \& {Freeman}(2016)}]{ness16}
{Ness} M, {Freeman} K. 2016.
\newblock \textit{\pasa} 33:e022

\bibitem[{{Nordlander} et~al.(2017){Nordlander}, {Amarsi}, {Lind}, {Asplund},
  {Barklem} et~al.}]{nordlander17}
{Nordlander} T, {Amarsi} AM, {Lind} K, {Asplund} M, {Barklem} PS, et~al. 2017.
\newblock \textit{\aap} 597:A6

\bibitem[{{Norris}, {Ryan} \& {Beers}(1997)}]{1997norriscarbon}
{Norris} JE, {Ryan} SG, {Beers} TC. 1997.
\newblock \textit{ApJ} 488:350

\bibitem[{{Payne}(1925)}]{payne25}
{Payne} CH. 1925.
\newblock \textit{{Stellar Atmospheres; a Contribution to the Observational
  Study of High Temperature in the Reversing Layers of Stars.}}
\newblock Ph.D. thesis, RADCLIFFE COLLEGE.

\bibitem[{{Pignatari} et~al.(2010){Pignatari}, {Gallino}, {Heil}, {Wiescher},
  {K{\"a}ppeler} et~al.}]{pignatari10}
{Pignatari} M, {Gallino} R, {Heil} M, {Wiescher} M, {K{\"a}ppeler} F, et~al.
  2010.
\newblock \textit{\apj} 710:1557

\bibitem[{{Pignatari} et~al.(2008){Pignatari}, {Gallino}, {Meynet}, {Hirschi},
  {Herwig} \& {Wiescher}}]{Pignatari08}
{Pignatari} M, {Gallino} R, {Meynet} G, {Hirschi} R, {Herwig} F, {Wiescher} M.
  2008.
\newblock \textit{\apjl} 687:L95--L98

\bibitem[{{Placco} et~al.(2013){Placco}, {Frebel}, {Beers}, {Karakas},
  {Kennedy} et~al.}]{placco13}
{Placco} VM, {Frebel} A, {Beers} TC, {Karakas} AI, {Kennedy} CR, et~al. 2013.
\newblock \textit{ApJ} 770:104

\bibitem[{{Placco} et~al.(2014){Placco}, {Frebel}, {Beers} \&
  {Stancliffe}}]{placco14}
{Placco} VM, {Frebel} A, {Beers} TC, {Stancliffe} RJ. 2014.
\newblock \textit{\apj} 797:21

\bibitem[{{Placco} et~al.(2017){Placco}, {Holmbeck}, {Frebel}, {Beers},
  {Surman} et~al.}]{Placco17}
{Placco} VM, {Holmbeck} EM, {Frebel} A, {Beers} TC, {Surman} RA, et~al. 2017.
\newblock \textit{\apj} 844:18

\bibitem[{{Planck Collaboration} et~al.(2016){Planck Collaboration}, {Ade},
  {Aghanim}, {Arnaud}, {Ashdown} et~al.}]{PlanckCosmology}
{Planck Collaboration}, {Ade} PAR, {Aghanim} N, {Arnaud} M, {Ashdown} M, et~al.
  2016.
\newblock \textit{\aap} 594:A13

\bibitem[{{Plez} et~al.(2004){Plez}, {Hill}, {Cayrel}, {Spite}, {Barbuy}
  et~al.}]{plez04}
{Plez} B, {Hill} V, {Cayrel} R, {Spite} M, {Barbuy} B, et~al. 2004.
\newblock \textit{A\&A} 428:L9

\bibitem[{{Qian} \& {Wasserburg}(2003)}]{qian_wasserburg03}
{Qian} YZ, {Wasserburg} GJ. 2003.
\newblock \textit{ApJ} 588:1099

\bibitem[{{Roederer}(2013)}]{Roederer13}
{Roederer} IU. 2013.
\newblock \textit{\aj} 145:26

\bibitem[{{Roederer} et~al.(2010){Roederer}, {Cowan}, {Karakas}, {Kratz},
  {Lugaro} et~al.}]{roederer10b}
{Roederer} IU, {Cowan} JJ, {Karakas} AI, {Kratz} KL, {Lugaro} M, et~al. 2010.
\newblock \textit{\apj} 724:975--993

\bibitem[{{Roederer} et~al.(2014){Roederer}, {Cowan}, {Preston}, {Shectman},
  {Sneden} \& {Thompson}}]{roederer14rproc_hbstars}
{Roederer} IU, {Cowan} JJ, {Preston} GW, {Shectman} SA, {Sneden} C, {Thompson}
  IB. 2014.
\newblock \textit{\mnras} 445:2970

\bibitem[{{Roederer} et~al.(2008{\natexlab{a}}){Roederer}, {Frebel},
  {Shetrone}, {Allende Prieto}, {Rhee} et~al.}]{cash1}
{Roederer} IU, {Frebel} A, {Shetrone} MD, {Allende Prieto} C, {Rhee} J, et~al.
  2008{\natexlab{a}}.
\newblock \textit{ApJ} 679:1549

\bibitem[{{Roederer} et~al.(2016{\natexlab{a}}){Roederer}, {Karakas},
  {Pignatari} \& {Herwig}}]{roederer16iproc}
{Roederer} IU, {Karakas} AI, {Pignatari} M, {Herwig} F. 2016{\natexlab{a}}.
\newblock \textit{\apj} 821:37

\bibitem[{{Roederer} et~al.(2009){Roederer}, {Kratz}, {Frebel}, {Christlieb},
  {Pfeiffer} et~al.}]{roederer09}
{Roederer} IU, {Kratz} K, {Frebel} A, {Christlieb} N, {Pfeiffer} B, et~al.
  2009.
\newblock \textit{ApJ} 698:1963

\bibitem[{{Roederer} et~al.(2012){Roederer}, {Lawler}, {Cowan}, {Beers},
  {Frebel} et~al.}]{roederer12b}
{Roederer} IU, {Lawler} JE, {Cowan} JJ, {Beers} TC, {Frebel} A, et~al. 2012.
\newblock \textit{\apjl} 747:L8

\bibitem[{{Roederer} et~al.(2008{\natexlab{b}}){Roederer}, {Lawler}, {Sneden},
  {Cowan}, {Sobeck} \& {Pilachowski}}]{roederer08iso}
{Roederer} IU, {Lawler} JE, {Sneden} C, {Cowan} JJ, {Sobeck} JS, {Pilachowski}
  CA. 2008{\natexlab{b}}.
\newblock \textit{\apj} 675:723

\bibitem[{{Roederer} et~al.(2016{\natexlab{b}}){Roederer}, {Mateo}, {Bailey},
  {Song}, {Bell} et~al.}]{Roederer16b}
{Roederer} IU, {Mateo} M, {Bailey} III JI, {Song} Y, {Bell} EF, et~al.
  2016{\natexlab{b}}.
\newblock \textit{\aj} 151:82

\bibitem[{{Rosswog} et~al.(1999){Rosswog}, {Liebend{\"o}rfer}, {Thielemann},
  {Davies}, {Benz} \& {Piran}}]{rosswog99}
{Rosswog} S, {Liebend{\"o}rfer} M, {Thielemann} FK, {Davies} MB, {Benz} W,
  {Piran} T. 1999.
\newblock \textit{\aap} 341:499

\bibitem[{{Sakari} et~al.(2018){Sakari}, {Placco}, {Hansen}, {Holmbeck},
  {Beers} et~al.}]{sakari18}
{Sakari} CM, {Placco} VM, {Hansen} T, {Holmbeck} EM, {Beers} TC, et~al. 2018.
\newblock \textit{\apjl} 854:L20

\bibitem[{{Schatz} et~al.(2002){Schatz}, {Toenjes}, {Pfeiffer}, {Beers},
  {Cowan} et~al.}]{schatz_chronometers}
{Schatz} H, {Toenjes} R, {Pfeiffer} B, {Beers} TC, {Cowan} JJ, et~al. 2002.
\newblock \textit{ApJ} 579:626

\bibitem[{{Shappee} et~al.(2017){Shappee}, {Simon}, {Drout}, {Piro}, {Morrell}
  et~al.}]{shappee17}
{Shappee} BJ, {Simon} JD, {Drout} MR, {Piro} AL, {Morrell} N, et~al. 2017.
\newblock \textit{arXiv:1710.05432}

\bibitem[{{Shetrone} et~al.(2003){Shetrone}, {Venn}, {Tolstoy}, {Primas},
  {Hill} \& {Kaufer}}]{shetrone03}
{Shetrone} M, {Venn} KA, {Tolstoy} E, {Primas} F, {Hill} V, {Kaufer} A. 2003.
\newblock \textit{\aj} 125:684--706

\bibitem[{{Simmerer} et~al.(2004){Simmerer}, {Sneden}, {Cowan}, {Collier},
  {Woolf} \& {Lawler}}]{simmerer04}
{Simmerer} J, {Sneden} C, {Cowan} JJ, {Collier} J, {Woolf} VM, {Lawler} JE.
  2004.
\newblock \textit{\apj} 617:1091--1114

\bibitem[{{Simon} et~al.(2015){Simon}, {Drlica-Wagner}, {Li}, {Nord}, {Geha}
  et~al.}]{simon15}
{Simon} JD, {Drlica-Wagner} A, {Li} TS, {Nord} B, {Geha} M, et~al. 2015.
\newblock \textit{\apj} 808:95

\bibitem[{{Sitnova}, {Mashonkina} \& {Ryabchikova}(2016)}]{sitnova16}
{Sitnova} TM, {Mashonkina} LI, {Ryabchikova} TA. 2016.
\newblock \textit{\mnras} 461:1000

\bibitem[{{Sneden}, {Cowan} \& {Gallino}(2008)}]{sneden08}
{Sneden} C, {Cowan} JJ, {Gallino} R. 2008.
\newblock \textit{\araa} 46:241--288

\bibitem[{{Sneden} et~al.(2000){Sneden}, {Cowan}, {Ivans}, {Fuller}, {Burles}
  et~al.}]{sneden2000}
{Sneden} C, {Cowan} JJ, {Ivans} II, {Fuller} GM, {Burles} S, et~al. 2000.
\newblock \textit{ApJ} 533:L139--L142

\bibitem[{{Starkenburg} et~al.(2017){Starkenburg}, {Oman}, {Navarro}, {Crain},
  {Fattahi} et~al.}]{starkenburg17}
{Starkenburg} E, {Oman} KA, {Navarro} JF, {Crain} RA, {Fattahi} A, et~al. 2017.
\newblock \textit{\mnras} 465:2212

\bibitem[{{Taylor} et~al.(2014){Taylor}, {Cinabro}, {Dilday}, {Galbany},
  {Gupta} et~al.}]{Taylor14}
{Taylor} M, {Cinabro} D, {Dilday} B, {Galbany} L, {Gupta} RR, et~al. 2014.
\newblock \textit{\apj} 792:135

\bibitem[{{Thielemann} et~al.(2017){Thielemann}, {Eichler}, {Panov} \&
  {Wehmeyer}}]{thielemann17}
{Thielemann} FK, {Eichler} M, {Panov} IV, {Wehmeyer} B. 2017.
\newblock \textit{Annual Review of Nuclear and Particle Science} 67:253

\bibitem[{{Travaglio} et~al.(2004){Travaglio}, {Gallino}, {Arnone}, {Cowan},
  {Jordan} \& {Sneden}}]{Travaglio04}
{Travaglio} C, {Gallino} R, {Arnone} E, {Cowan} J, {Jordan} F, {Sneden} C.
  2004.
\newblock \textit{\apj} 601:864--884

\bibitem[{{Tremblay} et~al.(2013){Tremblay}, {Ludwig}, {Freytag}, {Steffen} \&
  {Caffau}}]{tremblay13}
{Tremblay} PE, {Ludwig} HG, {Freytag} B, {Steffen} M, {Caffau} E. 2013.
\newblock \textit{\aap} 557:A7

\bibitem[{{Tsujimoto} et~al.(2017){Tsujimoto}, {Matsuno}, {Aoki}, {Ishigaki} \&
  {Shigeyama}}]{tsujimoto17}
{Tsujimoto} T, {Matsuno} T, {Aoki} W, {Ishigaki} MN, {Shigeyama} T. 2017.
\newblock \textit{\apjl} 850:L12

\bibitem[{{Tsujimoto} \& {Shigeyama}(2014)}]{Tsujimoto14b}
{Tsujimoto} T, {Shigeyama} T. 2014.
\newblock \textit{\apjl} 795:L18

\bibitem[{{Wallner} et~al.(2015){Wallner}, {Faestermann}, {Feige}, {Feldstein},
  {Knie} et~al.}]{Wallner15}
{Wallner} A, {Faestermann} T, {Feige} J, {Feldstein} C, {Knie} K, et~al. 2015.
\newblock \textit{Nature Communications} 6:5956

\bibitem[{{Wanajo}(2013)}]{Wanajo13}
{Wanajo} S. 2013.
\newblock \textit{\apjl} 770:L22

\bibitem[{{Wanajo} et~al.(2002){Wanajo}, {Itoh}, {Ishimaru}, {Nozawa} \&
  {Beers}}]{wanajo2002}
{Wanajo} S, {Itoh} N, {Ishimaru} Y, {Nozawa} S, {Beers} TC. 2002.
\newblock \textit{ApJ} 577:853

\bibitem[{{Wanajo} et~al.(2001){Wanajo}, {Kajino}, {Mathews} \&
  {Otsuki}}]{wanajo01}
{Wanajo} S, {Kajino} T, {Mathews} GJ, {Otsuki} K. 2001.
\newblock \textit{ApJ} 554:578

\bibitem[{{Wanajo} et~al.(2014){Wanajo}, {Sekiguchi}, {Nishimura}, {Kiuchi},
  {Kyutoku} \& {Shibata}}]{wanajo14}
{Wanajo} S, {Sekiguchi} Y, {Nishimura} N, {Kiuchi} K, {Kyutoku} K, {Shibata} M.
  2014.
\newblock \textit{\apjl} 789:L39

\bibitem[{{Winteler} et~al.(2012){Winteler}, {K{\"a}ppeli}, {Perego},
  {Arcones}, {Vasset} et~al.}]{Winteler12}
{Winteler} C, {K{\"a}ppeli} R, {Perego} A, {Arcones} A, {Vasset} N, et~al.
  2012.
\newblock \textit{\apjl} 750:L22

\bibitem[{{Wu} et~al.(2016){Wu}, {Fern{\'a}ndez}, {Mart{\'{\i}}nez-Pinedo} \&
  {Metzger}}]{Wu16}
{Wu} MR, {Fern{\'a}ndez} R, {Mart{\'{\i}}nez-Pinedo} G, {Metzger} BD. 2016.
\newblock \textit{\mnras} 463:2323--2334

\bibitem[{{Yoon} et~al.(2016){Yoon}, {Beers}, {Placco}, {Rasmussen}, {Carollo}
  et~al.}]{Yoon16}
{Yoon} J, {Beers} TC, {Placco} VM, {Rasmussen} KC, {Carollo} D, et~al. 2016.
\newblock \textit{arXiv:1607.06336}

\end{thebibliography}

\end{document}